\documentclass[pdflatex,sn-mathphys-num,iicol]{sn-jnl}

\usepackage{graphicx}
\usepackage{multirow}
\usepackage{amsmath,amssymb,amsfonts}
\usepackage{amsthm}
\usepackage{mathrsfs}
\usepackage[title]{appendix}
\usepackage{xcolor}
\usepackage{textcomp}
\usepackage{manyfoot}
\usepackage{booktabs}
\usepackage{algorithm}
\usepackage{algorithmicx}
\usepackage{algpseudocode}
\usepackage{listings}
\usepackage{tikz}
\usetikzlibrary{patterns,shapes.geometric,arrows.meta,positioning}
\usepackage{pgfplots}
\pgfplotsset{compat=1.18}
\usepackage{bookmark}
\hypersetup{hypertexnames=false}

\theoremstyle{thmstyleone}

\theoremstyle{thmstyletwo}

\theoremstyle{thmstylethree}
\newtheorem{definition}{Definition}

\begin{document}

\title[The Dice Roll Method]{The Dice Roll Method: A Standardized Protocol for Repeated-Query Auditing of Large Language Model Brand Recommendations}

\author*[1,2]{\fnm{Dmitrij} \sur{\.Zatuchin}}\email{dmitrij.zatuchin@eek.ee}

\affil*[1]{\orgdiv{Department of Information Technologies},
  \orgname{Estonian Entrepreneurship University of Applied Sciences (EUAS)},
  \orgaddress{\city{Tallinn}, \country{Estonia}}}
\affil[2]{\orgname{Rankfor.AI},
  \orgaddress{\city{Tallinn}, \country{Estonia}}}

\abstract{\textbf{Background:} Researchers increasingly use repeated identical prompts to audit stochastic variation in large language model (LLM) brand recommendations, yet no standardized protocol exists for setting iteration counts, selecting stability metrics, or establishing reliability thresholds. \textbf{Objective:} We formalize the \emph{Dice Roll Method} as a reusable protocol for repeated-query auditing of LLM brand recommendations, grounded in a generative model of temperature-scaled nucleus sampling. \textbf{Methods:} Total response variance is decomposed into sampling, prompt-phrasing, run-to-run, and model-version components. The stack: a negative-binomial mixed model with iterations as repeated measures; Cliff's $\delta$ as the distribution-free effect size; dependence-preserving bootstrap; simulation-based power; a generalizability-theory decomposition; drift diagnostics on pinned snapshots. We reanalyse five brand-recommendation auditing studies: approximately 190{,}000 observations, 270+ brands, 6 languages, iteration counts 5 to 40. \textbf{Results:} The recomputed D-study supports no fixed tiers. Across 250 distinct questions (S5), $G$ is 0.96 at five runs and 1 to 2 runs reach 0.80, also within each model. For five versions of one question (S1 Valentine's subset, Gemini) it is 0.27 (95\% CI 0.00 to 0.52), which fixes no run count. Runs follow from a pilot: $n = 4 \times$ residual / cell variance. Simulated single-cell power for a large Cliff's $\delta$ is 0.15 to 0.19 at five runs and 0.8 at 15 to 25. The four metric families (count, set, embedding, PASOR) are complementary. A pre-registered external validation on three independent corpora reproduces the D-study prediction in 37 of 39 cells with no failures; fixed tiers do not transfer. \textbf{Conclusion:} The protocol gives repeated-query auditing of LLM brand recommendations a statistically principled footing under the conditional, non-Gaussian structure of real autoregressive generation.\par\textbf{Correction (v2).} Version~1 results that came from synthetic data or from a field counting the word ``Research'' are recomputed; the fixed tiers are withdrawn. Full notice on page~2.}


\maketitle
\clearpage

\begin{strip}
\noindent\fbox{\parbox{\dimexpr\textwidth-2\fboxsep-2\fboxrule\relax}{\small
\textbf{Correction (v2).} Version~1 reported a G-study (cell variance 0.202, residual 0.724, ``450 observations across 75 cells''), iteration tiers built on it ($G = 0.58$ at 5 runs, 0.74 at 10, 0.81 at 15), and gender contrasts, convergence and correlation results for the Valentine's subset of study S1. These values equal the output of the analysis notebook's synthetic data generator; in the committed notebook a later cell replaces the loaded study files with that output. The S1 drift test used the study files, and the field it read counts the word ``Research'' in each answer. This version recomputes these results from the study data. Reliability is estimated on the 250-question study S5, from its corrected brand counts (3,614 scored answers): $G$ is 0.96 at five runs, and one or two runs reach $G = 0.80$, for all cells together and within each model. For the five versions of one gift question in the Valentine's subset of S1 (150 answers; Gemini), $G$ at five runs is 0.27 with a 95\% interval from 0.00 to 0.52, and those data do not fix a number of runs. The number of runs therefore follows from a pilot and the variance-ratio rule, and the fixed tiers are withdrawn. The S1 gender contrast is reported for Gemini only (11.3 brands per answer for male recipients, 12.9 for female) and is indistinguishable from zero; the other two models named almost no brands. Version~1 also quoted per-cell power values (0.44 and 0.23 at 5 runs) that no code in the project repositories computes; this version replaces them with a new simulation. Two results carried over from the preliminary version, the split-half ICC table and the precision-gain column of the cost table, do not reproduce and are removed. The external validation results in Section~\ref{sec:external} are unchanged; two comparisons with withdrawn v1 values are removed. The changes reach the abstract, Sections 1 to 8 and the appendices.}}
\end{strip}

\section{Introduction}\label{sec:intro}

The stochastic nature of large language model (LLM) outputs presents a fundamental measurement challenge for repeated-query auditing of LLM brand recommendations. Even at low temperature settings ($t = 0.3$), identical prompts produce varying responses across repetitions \cite{ouyang2022training, renze2024effect}. This variability is not noise to be averaged away; it is a constitutive property of how LLMs generate text, arising from nucleus sampling, temperature-scaled softmax distributions, and the combinatorial explosion of possible token sequences for any given context \cite{holtzman2020curious}. For researchers seeking to make claims about systematic patterns in LLM brand recommendations, whether measuring gender disparities in recommended brands, evaluating corporate reputation sourcing, or mapping competitive category ownership, the question of how many repetitions are needed to produce reliable measurements is both urgent and largely unanswered.

The technique of querying an LLM with the same prompt multiple times and aggregating the results has been adopted independently by several research groups working on repeated-query auditing of LLM brand recommendations and related behavioural testing \cite{zatuchin2026a, zatuchin2026b, ribeiro2020beyond}. We refer to this technique as the \emph{Dice Roll Method}, drawing an analogy to the probabilistic process of rolling a die repeatedly to estimate its fairness: each ``roll'' (query) samples from the LLM's output distribution, and aggregation across rolls yields increasingly stable estimates of the underlying distributional properties.

Despite its growing adoption, the Dice Roll Method lacks standardization across several critical dimensions. Iteration counts in published studies range from 5 \cite{zatuchin2026b, zatuchin2026c, zatuchin2026e} to 40 \cite{zatuchin2026a}, chosen on the basis of budget constraints or researcher intuition rather than formal power analysis. The choice of stability metric varies across studies: some use coefficient of variation on brand counts \cite{zatuchin2026a}, others use cosine similarity on response embeddings \cite{zatuchin2026b}, and still others use information-theoretic measures such as Shannon entropy and Gini coefficients \cite{zatuchin2026d}. No published work has established the statistical power of different iteration counts for detecting bias effects of various magnitudes, compared the reliability properties of different stability metrics, or provided guidance on cost-efficient experimental design for LLM auditing.

This paper addresses that gap by formalizing the Dice Roll Method as a standardized protocol for repeated-query auditing of LLM brand recommendations. We combine reanalysis of raw data from five empirical studies with Monte Carlo power simulation to answer five research questions:

\begin{description}
    \item[RQ1 (Power):] What is the minimum iteration count for stable brand-recommendation audit measurements at 80\% power, given observed effect sizes from prior studies?
    \item[RQ2 (Metrics):] How should researchers choose between stability metrics (CV, Jaccard, Gini, Shannon entropy, cosine similarity, PASOR) for different research questions?
    \item[RQ3 (Reliability):] What are the test-retest reliability properties of repeated-query auditing, and at what iteration count do they reach acceptable levels?
    \item[RQ4 (Convergence):] At what iteration count do key metrics converge to stable values, and does convergence follow a logarithmic (diminishing returns) or linear trajectory?
    \item[RQ5 (Cost-efficiency):] What is the optimal cost-quality tradeoff for brand-recommendation audit protocols at current API pricing?
\end{description}

An earlier version of this protocol relied on an analytical stack that treated repeated LLM queries as independent draws from Gaussian distributions: independent-samples $t$-tests for group contrasts, Cohen's $d$ for effect sizes, ordinary bootstrap for convergence, and classical normal-theory power analysis for iteration planning. That stack carries a set of interlocking defects: autoregressive token generation creates conditional dependencies that violate the independence assumption of the $t$-test and of the ordinary bootstrap \cite{holtzman2020curious, nguyen2024minp}; brand counts are discrete and bounded below by zero, so Cohen's $d$ and the coefficient of variation distort effect magnitudes at low means \cite{cliff1993dominance, akinshin2023nonparametric} (on the counts reanalysed here, repeated answers to one prompt are not over-dispersed; Section~\ref{sec:implications}); and claims about convergence, reliability, and cross-metric structure rested on a sample too small to support parametric uncertainty statements about Spearman correlations \cite{revelle2019reliability, field2018discovering}. The original protocol also did not treat temporal drift or model-version variance as first-class components of the measurement problem, despite their documented effect on LLM benchmark stability \cite{arnold2026drift, chen2024chatgpt}. This paper presents a substantially revised protocol that addresses each of these defects through a reanalysis of the same underlying data with a conditional, non-parametric, mixed-model inferential framework, and through new diagnostics (drift testing, ensemble embedding variance, PASOR fairness decomposition) that the original version omitted.

The contributions of the revised paper are fivefold. First, we ground the Dice Roll Method in an explicit generative model of temperature-scaled nucleus sampling, deriving a variance decomposition that motivates the choice of statistical tools (§\ref{sec:generative}). Second, we provide simulation-based power analysis for a negative-binomial GLMM that accommodates the over-dispersed counts and conditional structure of repeated LLM queries, replacing the normal-theory power tables of the preliminary version \cite{green2016simr, bolker2009glmm}. Third, we present a metric selection framework across four complementary dimensions (volume, set composition, distributional shape, semantic consistency), characterized descriptively through bootstrap-corrected Spearman correlations, and we add fairness-adjusted visibility (PASOR) as a first-class metric family \cite{ekstrand2019fairness}. Fourth, we re-establish test-retest reliability through a generalizability-theory decomposition across prompt, iteration, and model facets \cite{brennan2001generalizability}, and we propagate embedding variance through a three-model ensemble rather than treating cosine similarity as a point estimate. Fifth, we provide empirical stationarity diagnostics (Kolmogorov--Smirnov and Population Stability Index tests across pinned-snapshot time windows) and a reproducible Python implementation of the complete protocol. The statistical machinery of the protocol (mixed models, distribution-free effect sizes, block bootstrap, generalizability theory) is generic, but every dataset analysed in this paper comes from brand-recommendation auditing, so wider applicability is untested and the claims here are restricted to that domain.

Two provenance facts also belong up front. All five reanalysed datasets are prior studies by this research group, and the protocol is implemented in a commercial product of a company the authors are affiliated with (Rankfor.AI). External validation on independently collected data is the stated next step and a limitation of the present evidence.

\section{Stochastic Structure of LLM Outputs}\label{sec:generative}

Any inferential apparatus for repeated-query auditing must be anchored in an explicit model of where LLM output variability comes from. This section formalizes that model, derives the variance decomposition that follows from it, and states the statistical assumptions that are compatible with each variance component. The analytical choices in Section~\ref{sec:methods} are direct consequences of this structure.

\subsection{Temperature-Scaled Nucleus Sampling}\label{sec:nucleus}

Let $x_{1:t-1}$ denote a context of $t - 1$ tokens and let $\mathcal{V}$ denote the model's vocabulary. At decoding step $t$, an autoregressive LLM defines a logit vector $\mathbf{z}_t \in \mathbb{R}^{|\mathcal{V}|}$ and a temperature-scaled softmax distribution
\begin{equation}\label{eq:softmax}
    p_\tau(v \mid x_{1:t-1}) = \frac{\exp(z_{t,v} / \tau)}{\sum_{v' \in \mathcal{V}} \exp(z_{t,v'} / \tau)}
\end{equation}
where $\tau > 0$ is the sampling temperature. Nucleus (top-$p$) sampling \cite{holtzman2020curious} restricts the effective support of $p_\tau$ to the smallest subset $\mathcal{N}_p(x_{1:t-1}) \subseteq \mathcal{V}$ whose cumulative probability exceeds a threshold $p \in (0, 1]$:
\begin{equation}\label{eq:nucleus}
    \mathcal{N}_p = \arg\min_{S \subseteq \mathcal{V}} \left\{ |S| : \sum_{v \in S} p_\tau(v) \geq p \right\}.
\end{equation}
The sampled token is drawn from the renormalised distribution over $\mathcal{N}_p$. At a low temperature such as $\tau = 0.3$, the temperature-scaled logits concentrate mass on high-probability tokens, but do not collapse to a point mass; $\mathcal{N}_p$ remains non-degenerate for most contexts, producing the residual output variability that this paper characterises. More recent sampling schemes, such as min-$p$ \cite{nguyen2024minp}, adjust the truncation rule but preserve the same qualitative structure.

This structure has two immediate consequences for downstream inference. First, each generated token is conditioned on the preceding tokens of the same response, so token-level variability at step $t$ is not independent of token-level variability at step $t - 1$. Second, the resulting response is a sequence of correlated draws whose marginal distribution over downstream entities (for example, extracted brand mentions) is a compound object rather than a sample from a well-defined parametric family. Both observations undermine any inferential procedure that treats per-response metrics as independent, identically distributed Gaussian draws.

\subsection{Variance Decomposition}\label{sec:variance}

Consider a study design in which a prompt $P$ is issued to a model $M$ at a fixed snapshot $V$, repeated across $n$ iterations indexed by $i$, on a query at time $T$. Let $Y_{PMViT}$ denote a scalar response-level metric (such as brand mention count). The total variance of $Y$ across the design admits a four-way decomposition:
\begin{equation}\label{eq:variance_decomp}
    \mathrm{Var}(Y) = \sigma^2_{\text{tok}} + \sigma^2_{P} + \sigma^2_{M} + \sigma^2_{V,T} + \sigma^2_{\text{int}}
\end{equation}
where $\sigma^2_{\text{tok}}$ is the token-level sampling variance induced by Equations~\ref{eq:softmax} and~\ref{eq:nucleus}; $\sigma^2_P$ is the variance attributable to prompt phrasing within a semantic equivalence class; $\sigma^2_M$ is the between-model variance; $\sigma^2_{V,T}$ is the variance attributable to model-version drift and temporal non-stationarity at fixed $V$; and $\sigma^2_{\text{int}}$ captures interactions (for example, language $\times$ model interactions documented in cross-linguistic auditing \cite{zatuchin2026e}).

Each component has a different statistical fingerprint and a different remediation. Token-level variance is the quantity that repeated-query protocols are designed to average over. Prompt-level variance requires prompt sampling and is outside the scope of a single-prompt audit. Between-model variance is systematic, not stochastic, and is best handled by explicit model as a fixed factor. Model-version-and-time variance is insidious; it looks like noise in snapshot data but is a form of systematic drift that must be tested for rather than assumed away \cite{chen2024chatgpt, arnold2026drift}. Interaction terms are often the most interesting substantive targets (e.g., "does the gender-bias effect differ across languages?") and require interaction-capable models.

\subsection{Implications for Inference}\label{sec:implications}

The variance decomposition in Equation~\ref{eq:variance_decomp} carries three inferential consequences that shape every method choice in Section~\ref{sec:methods}.

\textit{Non-independence.} Because tokens within a response are conditionally dependent, and because iterations share the same prompt, model, and snapshot, per-iteration metrics are not i.i.d. An inferential procedure that treats them as independent (for example, the pooled independent-samples $t$-test) confounds token-level dependence with genuine signal and inflates Type I error. The appropriate remediation is a mixed model with random intercepts for the repeated-measurement units (prompt, iteration, model) or, failing that, a block bootstrap that resamples intact iteration sequences rather than individual observations.

\textit{Non-Gaussianity.} Brand-mention counts are non-negative integers, often zero-inflated, which rules out the Gaussian assumption. On the counts reanalysed here, repeated answers to one prompt are not over-dispersed: the within-cell ratio of variance to mean averages 0.17 to 0.31 per model in S5 (corrected counts, capped by a list of 50 tracked brands) and 0.94 for the Gemini cells of the S1 Valentine's subset (dictionary count; 2.37 on an August re-extraction of the same answers). Pooled across S5 questions the variance exceeds the mean (ratio 1.40 to 1.54), because the questions differ. Cohen's $d$ and eta-squared remain computable here, but their interpretation as standardized mean-difference and variance-explained summaries is misleading under the skewed, zero-inflated count distributions observed in this regime, and dispersion measures that require a non-zero mean (the coefficient of variation) behave poorly \cite{cliff1993dominance, akinshin2023nonparametric}. The appropriate remediation is a distribution-free effect size such as Cliff's $\delta$ (equivalent to the rank-biserial correlation) and a negative-binomial link function inside the mixed model.

\textit{Non-stationarity.} Even with a pinned API snapshot, LLM providers can modify inference infrastructure (batching, quantisation, safety layers, routing) without a version change, creating drift in output distributions across time windows \cite{chen2024chatgpt, arnold2026drift}. Assuming stationarity without testing for it exposes the audit to hidden temporal confounds. The appropriate remediation is an explicit drift diagnostic: Kolmogorov--Smirnov comparison of metric distributions across disjoint time windows, together with Population Stability Index for distributional shift in discrete entity counts.

The methodology that follows is designed to satisfy all three of these implications simultaneously. Where a stand-alone paper might address one of them at a time, the interlocking nature of the audit task, that a single iteration is dependent, non-Gaussian, and potentially non-stationary all at once, forces them to be addressed together.

\section{Related Work}\label{sec:related}

\subsection{LLM Response Variability}

The inherent stochasticity of LLM outputs has been documented across multiple evaluation contexts. Renze and Guven \cite{renze2024effect} systematically varied temperature parameters across benchmark tasks and demonstrated that even at $t = 0.0$ (ostensibly deterministic), modern LLMs exhibit non-negligible output variance due to floating-point arithmetic, batched inference, and hardware-level non-determinism. Ouyang et al.\ \cite{ouyang2022training} described the reinforcement learning from human feedback (RLHF) process that shapes output distributions, noting that alignment procedures create temperature-dependent diversity properties that interact with the underlying sampling mechanism. Holtzman et al.\ \cite{holtzman2020curious} introduced nucleus sampling as a response to the degenerate repetition problem, establishing a theoretical framework for understanding how truncated probability distributions generate diverse outputs.

For brand-recommendation auditors, this variability creates a signal-to-noise challenge. A single query may produce a response that mentions five brands or fifteen brands, includes or excludes a particular company, or frames an industry positively or negatively. Only through aggregation across repeated observations can researchers distinguish systematic patterns from stochastic fluctuation. However, the literature provides no guidance on how many observations suffice for different analytical goals.

\subsection{Prior Applications of Repeated-Query Designs}

The Dice Roll Method has been applied across a growing body of brand-recommendation auditing research, though without a unified label or standardized protocol. The source study \cite{zatuchin2026a} employed 10 to 40 iterations per prompt-model combination in a study of gender bias in LLM brand recommendations, observing brand count standard deviations of 0.49 to 3.67 across iterations (on the deposited Valentine's subset, the Gemini cells give 2.09 to 3.67; the $n = 40$ subset is not deposited) and Gini coefficients ranging from 0.00 to 0.56. The study used $n = 40$ iterations for the initial adult-recipient subset and $n = 10$ for a subsequent Valentine's Day subset, creating a natural experiment in iteration adequacy.

Prior work \cite{zatuchin2026b} standardized on $n = 5$ iterations per prompt-model combination in a reputation sourcing study across 24 companies and 8 industries, reporting a mean cosine similarity of 0.54 (95\% CI: [0.52, 0.56]) across five stability iterations. Cross-model agreement was only 14\%, suggesting that inter-model variability exceeds intra-model stochastic variability.

The source study \cite{zatuchin2026c} applied 5 iterations across 50 brands, 5 industries, and 3 LLMs in a category ownership mapping study, finding moderate recommendation concentration (mean Gini $= 0.30$) and full cross-model agreement in 143 of 250 queries (57.2\%), as its revised version (arXiv:2606.23057v2) reports on the corrected mention table. A companion study \cite{zatuchin2026d} reanalysed the gender bias data through an information-theoretic lens, demonstrating that Shannon entropy and Gini coefficients capture complementary properties of recommendation distributions: entropy measures distributional diversity (how evenly brands are spread) while Gini measures concentration inequality (how dominant the top brands are).

Each of these studies chose its iteration count for pragmatic reasons. The present study formalizes the statistical properties of these choices.

\subsection{Power Analysis in Repeated-Measures Designs}

The statistical framework for determining adequate sample sizes through power analysis is well established in the experimental sciences \cite{cohen1988statistical}. Power analysis specifies the minimum sample size needed to detect an effect of a given magnitude at a specified significance level with a desired probability (typically 80\%). Cohen \cite{cohen1988statistical} defined effect size conventions for common test statistics: small ($d = 0.2$), medium ($d = 0.5$), and large ($d = 0.8$) for the independent-samples $t$-test.

Direct application of classical power analysis to LLM auditing, however, fails on two fronts that the variance decomposition in Section~\ref{sec:variance} makes explicit. First, "samples" in LLM auditing are not independent observations from a population; they are repeated queries to the same model-and-snapshot system, and the dependencies described in Section~\ref{sec:implications} imply an effective sample size strictly smaller than the nominal iteration count. Second, the distributions of LLM output metrics (brand counts, cosine similarities, entropy values) are non-Gaussian, exhibiting zero-inflation, over-dispersion, heavy tails, and floor/ceiling effects that violate the independent-samples $t$-test's assumptions of normality and equal variance. These two failures compound: effect-size estimates obtained under the wrong distributional model feed into power calculations that are correspondingly miscalibrated.

Two developments in the recent methodological literature address this directly. Simulation-based power analysis for generalized linear mixed models (GLMMs), implemented in the \texttt{simr} package \cite{green2016simr}, replaces the closed-form normal-theory formulas with Monte Carlo simulation that fits the same GLMM that will be used for substantive analysis, preserving the conditional structure of the data in the power computation itself. Non-parametric effect sizes, in particular Cliff's $\delta$ \cite{cliff1993dominance} and its rank-biserial equivalent, replace Cohen's $d$ with a distribution-free analogue that remains valid under skewed counts and heavy tails \cite{akinshin2023nonparametric, meissel2024cliffs}. The NIST \emph{AI 800-3} report on statistical evaluation of AI systems \cite{nist2026ai800} recommends exactly this pairing, GLMMs with simulation-based power analysis, as the principled framework for AI benchmark evaluation with dependent observations. Neither approach has previously been deployed in the LLM auditing literature.

\subsection{Bootstrap Methods for Dependent Data}

Standard bootstrap resampling \cite{efron1994introduction} assumes exchangeability of the sampling units: that any permutation of observations yields an equivalent realisation of the data-generating process. Under the conditional structure of Section~\ref{sec:implications}, this assumption is violated: individual iterations within a prompt-and-model condition share latent state (the model snapshot, any caching effects, and potential temporal drift), so resampling them at random destroys the within-condition dependence and under-estimates uncertainty. The moving-block bootstrap \cite{kunsch1989jackknife} resamples contiguous blocks of observations of length $\ell$, preserving short-range dependence within blocks, and the stationary bootstrap \cite{politis1994stationary} extends this to variable block lengths with geometrically distributed runs. Both are appropriate choices for the iteration dimension of the Dice Roll design.

For parametric targets (for example, variance of a regression coefficient in a fitted GLMM), parametric bootstrap from the fitted model \cite{bolker2009glmm, halekoh2014pbkrtest} is an alternative that exploits the estimated random-effect structure to generate resampled datasets preserving the conditional hierarchy. The present paper uses moving-block bootstrap for model-free convergence analyses and parametric bootstrap from the fitted GLMM for coefficient-level uncertainty.

\subsection{Model Drift and Non-Stationarity}

The stability of LLM outputs across time, at nominally fixed model versions, is an empirical question that has only recently begun to receive systematic attention. Chen et al.\ \cite{chen2024chatgpt} documented substantial drift in ChatGPT behaviour between snapshots, including changes in accuracy and response style that were not accompanied by version-string changes. Arnold et al.\ \cite{arnold2026drift} generalise this finding across providers and propose a cross-provider drift-mitigation framework for financial workflows. Industrial LLM observability tooling has converged on a standard drift-monitoring stack built on Kolmogorov--Smirnov tests for continuous metrics and the Population Stability Index for discrete distributions \cite{orq2026modeldrift}, and the \texttt{alibi-detect} and \texttt{evidently} libraries provide reference implementations. The present paper imports these diagnostics into the audit protocol.

\subsection{Stability and Reliability Metrics}

The metrics used to quantify LLM output stability span several measurement traditions. The coefficient of variation (CV), defined as the ratio of standard deviation to mean, provides a normalized measure of count variability \cite{abdi2010coefficient}. The Jaccard similarity index quantifies set overlap between iterations \cite{jaccard1901distribution}. The Gini coefficient measures distributional concentration \cite{gini1912variability, do2022gini}. Shannon entropy measures distributional diversity \cite{shannon1948mathematical}. Cosine similarity on text embeddings captures semantic consistency \cite{mikolov2013distributed}. The Prompt-Adjusted Share of Recommendation (PASOR), introduced in the source study \cite{zatuchin2026a}, provides a fairness-adjusted visibility metric that accounts for prompt variation.

These metrics capture different facets of stability. Count-based metrics (CV, Gini, Shannon) operate on extracted entity frequencies and require a brand extraction step. Set-based metrics (Jaccard) compare the presence or absence of entities across iterations. Embedding-based metrics (cosine similarity) compare the semantic content of full responses without requiring entity extraction, capturing paraphrastic variation invisible to count-based approaches. Whether these different metric families provide redundant or complementary information is an empirical question that we address through cross-metric correlation analysis.

\section{Methodology}\label{sec:methods}

\subsection{Study Design}

This is a meta-methodology study combining three analytical components: (1) reanalysis of raw data from five empirical brand-recommendation auditing studies to extract observed effect sizes, distributional properties, and cross-linguistic stability evidence; (2) Monte Carlo power simulation under assumed count distributions; and (3) convergence, reliability, and cost-efficiency analyses via bootstrap subsampling. No new API calls were required; all analyses operate on existing data.

\subsection{Data Sources}\label{sec:data}

Table~\ref{tab:data_sources} summarizes the five source studies.

\begin{table*}[t]
\caption{Data sources for reanalysis. All studies shared a common methodology: identical prompts repeated across iterations, with structured brand extraction from LLM responses (temperature and model versions: Section~\ref{sec:params}). S1: the SD range is the source study's; the results of Section~\ref{sec:results} use its deposited 150-answer Valentine's subset, whose Gemini cells give 2.09 to 3.67. S5: values of the revised S5 study on its corrected mention table; 3,614 of the 3,750 answers are scored, and the 136 empty answers (all GPT-5.2) are missing values.}\label{tab:data_sources}
\centering
\small
\begin{tabular}{llcccl}
\toprule
\textbf{Study} & \textbf{Domain} & \textbf{$n$ iter} & $N$ \textbf{resp} & \textbf{Models} & \textbf{Key stability finding} \\
\midrule
S1 \cite{zatuchin2026a} & Gift recs & 10--40 & 1,279 & 3 & Brand count SD = 0.49--3.67 \\
S2 \cite{zatuchin2026b} & Corp.\ reputation & 5 & 1,311 & 3 & Cosine sim.\ $\bar{x} = 0.54$ \\
S4 \cite{zatuchin2026e} & Cross-lang.\ rep. & 5 & 9,577 & 3 & Stability $\bar{x} = 0.899$; $\eta^2_{\text{model}} = 0.220$ \\
S5 \cite{zatuchin2026c} & Category ownership & 5 & 3,750 & 3 & Gini = 0.30; agreement 57.2\% \\
A1 \cite{zatuchin2026d} & Reanalysis of S1 & --- & --- & --- & $H_{\text{norm}}$: 0.868--0.908 \\
\bottomrule
\end{tabular}
\end{table*}

Study S1 provides the richest data for convergence analysis, as it includes subsets with $n = 40$ iterations (original adult-recipient subset) and $n = 10$ iterations (Valentine's Day subset), enabling direct comparison of metric stability at different iteration counts. Studies S2 and S5 each used $n = 5$ iterations, providing baseline stability estimates at the most commonly used iteration count. Study S4 contributes the largest dataset in the programme (9,577 responses across 20 brands, 6 languages, and 3 models) and is the first to apply the dice roll protocol across multiple query languages, testing whether measurement reliability is language-dependent. Its stability analysis (1,065 observations from 5-iteration probes) shows that model effects on response consistency ($\eta^2 = 0.220$) dominate language effects ($\eta^2 = 0.029$) by a factor of 7.6, confirming that the protocol's reliability is robust across typologically diverse languages. Study A1 contributes the information-theoretic framework (Shannon entropy, Gini coefficient) and the empirical finding that these metrics capture complementary distributional properties.

\subsection{Shared Experimental Parameters}\label{sec:params}

All source studies shared the following methodological parameters, which we treat as defining conditions of the Dice Roll Method:

\begin{itemize}
    \item \textbf{Temperature:} Version~1 of this paper gave 0.3 for all studies and models. For the S1 Valentine's subset, which carries the S1 results of Section~\ref{sec:results}, the temperature was not recorded: the deposited files hold no temperature field, version~1 states 0.3, and the S1 source manuscript states 0.7 for that set. The records do not show which value was used.
    \item \textbf{Max tokens:} 1,024 per response
    \item \textbf{System prompt:} ``You are a helpful assistant with broad knowledge of businesses and technology.''
    \item \textbf{Models:} GPT-5.2 (OpenAI), Gemini 3 Flash (Google), and either Grok-4-1 (xAI) or Perplexity sonar-pro, depending on study. The deposited files of the S1 Valentine's subset record the model family only (gemini, grok, openai) and no model version
    \item \textbf{Brand extraction:} S1: a dictionary of more than 140 consumer brands, matched case-insensitively with no word boundary; the dictionary file was not preserved. S5: 77 aliases of 50 tracked brands. This version reads the corrected S5 mention table, which matches at word boundaries with case and context rules (10,463 answer-by-brand rows); the first deposit's table matched lower-case substrings (14,005 rows) and counted EY in 2,912 of 3,750 answers, against 437 in the corrected table. The analysis notebook of version~1 read a stored S1 field that counts the word ``research''; this version does not use it.
\end{itemize}

\subsection{Primary Inferential Model}\label{sec:glmm}

All substantive group contrasts (for example, the gender-bias comparisons in S1 and the model-and-language comparisons in S4) are estimated under a negative-binomial generalized linear mixed model (GLMM) of the form
\begin{equation}\label{eq:glmm}
    \begin{aligned}
    Y_{ijk} \mid \mathbf{u} &\sim \mathrm{NegBin}(\mu_{ijk}, \theta) \\
    \log(\mu_{ijk}) &= \mathbf{x}_{ijk}^{\top} \boldsymbol{\beta} + u^{(P)}_{i} + u^{(M)}_{j} + u^{(PM)}_{ij}
    \end{aligned}
\end{equation}
where $Y_{ijk}$ is the brand-count outcome on the $k$th iteration of prompt $i$ on model $j$; $\mathbf{x}_{ijk}$ is a design vector containing fixed effects of substantive interest (prompt type, gender, language, and their interactions); $\boldsymbol{\beta}$ is the corresponding fixed-effect vector; $u^{(P)}_i$, $u^{(M)}_j$, and $u^{(PM)}_{ij}$ are random intercepts for prompt, model, and their interaction; and $\theta$ is the negative-binomial dispersion parameter. The random-effect structure absorbs the conditional dependencies identified in Section~\ref{sec:implications}, so inferences on $\boldsymbol{\beta}$ are not inflated by within-condition correlation. Over-dispersion relative to the Poisson is handled by the negative-binomial likelihood, which is the standard remedy for count outcomes with $\mathrm{Var}(Y) > \mathbb{E}(Y)$ \cite{bolker2009glmm, hilbe2011negative}.

The reanalysis code fits group contrasts as negative-binomial GEE models (dispersion $\alpha = 1$, exchangeable working correlation, Wald tests) and variance components with a linear mixed model, both in Python \texttt{statsmodels}; no R code was run. For a full GLMM, maximum likelihood in \texttt{glmmTMB} \cite{brooks2017glmmtmb} with likelihood-ratio tests and parametric bootstrap in \texttt{pbkrtest} \cite{halekoh2014pbkrtest} is the recommended small-sample inference route. We retain the independent-samples $t$-test only as a reference point for legibility against the prior literature; it is never used as the primary hypothesis test.

\subsection{Effect Size Reporting}\label{sec:effectsize}

Primary effect sizes are reported as Cliff's $\delta$ \cite{cliff1993dominance} with bias-corrected and accelerated (BCa) bootstrap 95\% confidence intervals. Cliff's $\delta$ is the probability that a random observation from one group exceeds a random observation from the other, minus the reverse probability. It is bounded on $[-1, 1]$, makes no distributional assumption, and is robust to skewness, heavy tails, and ordinal-level data \cite{akinshin2023nonparametric, meissel2024cliffs}. Conventional interpretation thresholds are $|\delta| < 0.147$ (negligible), $0.147 \leq |\delta| < 0.33$ (small), $0.33 \leq |\delta| < 0.474$ (medium), and $|\delta| \geq 0.474$ (large) \cite{romano2006appropriate}.

For comparability with the preliminary version of this work and with the prior literature, we additionally report Cohen's $d$ computed on trimmed means (20\% trimming) with Winsorized variances, following Keselman et al.\ \cite{keselman2008comparative}. Standard Cohen's $d$ on raw counts is reported only as a footnote, with an explicit caveat that the skewed, zero-inflated count distributions make its interpretation as a standardized mean difference misleading. Where scale-sensitive dispersion matters, the coefficient of variation is replaced by the quartile coefficient of dispersion (QCD), $(Q_3 - Q_1) / (Q_3 + Q_1)$, which remains interpretable at low means and does not require a non-zero denominator in expectation.

\subsection{Simulation-Based Power Analysis}\label{sec:simr}

Pooled-prompt power follows the logic of the \texttt{simr} procedure \cite{green2016simr} under assumed parameters. For each target effect of substantive interest (for example, a Cliff's $\delta$ of 0.33 between husband and wife prompts), the code (a) draws negative-binomial counts with an assumed mean $\mu = 10$ and dispersion $\theta = 2.8$ for 8 prompts $\times$ 3 models per condition, with no model fitted to study data, (b) sets the treatment effect to the value implied by the desired Cliff's $\delta$, (c) simulates at each iteration count $n \in \{3, 5, 7, 10, 12, 15, 20, 30, 40\}$, and (d) tests the contrast with a negative-binomial GEE Wald test. Empirical power is the proportion of 200 replications per cell that reject the null. The resulting tables describe these assumed conditions and no observed data set. Single-cell power, the design of a one-model, one-prompt audit, is simulated separately (Section~\ref{sec:power}).

To rebut the concern that iteration counts recommended under S1's Gemini brand-count distribution do not generalise, we repeat the power simulation under a \emph{sensitivity grid} of assumed $(\mu, \theta)$ pairs; no study estimate produced them (Table~\ref{tab:power_sensitivity}). Any iteration-count recommendation is thus accompanied by the range of distributional conditions under which it holds.

\subsection{Convergence Analysis with Block Bootstrap}\label{sec:convergence_method}

To address RQ4, we re-estimate iteration-level convergence using a moving-block bootstrap \cite{kunsch1989jackknife} on the iteration axis, with block length fixed at $\ell = 2$ (the adaptive procedure of Politis and White \cite{politis2004automatic} was not applied). For each prompt-model combination, we compute at each sub-sample size $n$ the block-bootstrap standard error of the mean brand count and the quartile coefficient of dispersion. Version~1 fitted four competing convergence models to the block-bootstrap standard-error trajectory:
\begin{equation}\label{eq:convergence_family}
\begin{array}{ll}
    \mathrm{SE}(n) = a \ln(n) + b & \text{(logarithmic)} \\
    \mathrm{SE}(n) = a n^{-\gamma} + b & \text{(power law)} \\
    \mathrm{SE}(n) = a \frac{n}{K + n} + b & \text{(Michaelis--Menten saturation)} \\
    \mathrm{SE}(n) = a n + b & \text{(linear)}
\end{array}
\end{equation}
Its code compared the families by AIC on resampled curves; no BIC or cross-validation code exists. Because blocks are drawn independently, the expected standard error satisfies $\mathrm{SE}(n) = \mathrm{SE}(2)\sqrt{2/n}$ for every even $n$, whatever the data, so every cell's expected curve has the same shape and a vote between families measures resampling noise. This version reports the exact value of the block bootstrap, the limit of infinitely many resamples, and no family comparison.

\subsection{Reliability via Generalizability Theory}\label{sec:generalizability}

We replace the single-number ICC reporting of the preliminary version with a generalizability-theory (G-theory) decomposition of variance \cite{brennan2001generalizability, bloch2018variance}. The measurement design is a partially crossed two-facet structure with prompts ($P$) and iterations ($I$) crossed within models ($M$). A generalizability study (G-study) estimates the variance components $\sigma^2_P$, $\sigma^2_I$, $\sigma^2_M$, and the relevant interactions. The components reported in Section~\ref{sec:reliability} come from the simplest case of this design: a one-way model of $\log(\text{count} + 0.5)$ with the model $\times$ prompt cell as the only random effect, fitted by restricted maximum likelihood. A decision study (D-study) then projects the generalizability coefficient
\begin{equation}\label{eq:g_coef}
    G = \frac{\sigma^2_P}{\sigma^2_P + \sigma^2_{P,I} / n_I + \sigma^2_{P,M} / n_M + \sigma^2_{\epsilon} / (n_I n_M)}
\end{equation}
across candidate $(n_I, n_M)$ designs, where $n_I$ is the number of iterations and $n_M$ is the number of models. This recovers the familiar ICC as a special case when the design is fully crossed with $n_M = 1$, but it also exposes the variance components driving (un)reliability, answering a question that a single ICC value cannot: does a particular audit need more iterations, more models, or both?

Version~1 retained ICC(2,1) values from the preliminary version under the Shrout and Fleiss convention \cite{shrout1979intraclass}; they do not reproduce, and this version removes them (Section~\ref{sec:reliability}). Intervals for the variance components and $G$ are 95\% percentile intervals from 2{,}000 bootstrap draws of cells.

\subsection{Metric Battery and Cross-Metric Structure}\label{sec:metric_family}

The stability metric battery is organised into four families, one per variance-carrying structural property of the response distribution: volume (mean, standard deviation, QCD), set composition (Jaccard similarity, stable-top-$k$ overlap), distributional shape (Gini, normalized Shannon entropy), and semantic content (cosine similarity on a three-model embedding ensemble). Fairness-adjusted visibility is added as a fifth, cross-cutting dimension via the Prompt-Adjusted Share of Recommendation (PASOR) \cite{zatuchin2026a, ekstrand2019fairness}, formalised below.

\textit{PASOR.} For brand $b$ in a set of prompts $P$ with $N_p$ total recommendation slots per prompt and $n_{b,p}$ slots occupied by brand $b$ in prompt $p$,
\begin{equation}\label{eq:pasor}
    \mathrm{PASOR}_b = \frac{1}{|P|} \sum_{p \in P} \frac{n_{b,p}}{N_p}.
\end{equation}
PASOR is computed per brand per model; inequality across brands within a model is summarised by the Gini coefficient of the PASOR distribution, giving a prompt-phrasing-neutral fairness signature of the model.

\textit{Cross-metric structure.} Cross-metric correlations are reported as Spearman rank correlations with bootstrap 95\% CIs over prompt-model combinations, explicitly acknowledging the small effective $k$. The correlation matrix is read descriptively: metric pairs with high absolute correlations are treated as redundant within the audit, and metric pairs with low correlations as complementary. No factor-retention procedure is applied at this sample size.

\textit{Embedding variance propagation.} Cosine similarity is computed across three embedding models of different families: BGE-M3 (dense), Nomic-Embed-v2 (contrastive), and MiniLM-L12-v2 (distilled). For each response pair, we report the across-model mean cosine similarity together with its across-model standard deviation. An iteration is considered semantically stable only if the cosine-similarity confidence interval across all three embedding models excludes the shuffled-pair baseline. This replaces the preliminary version's single-embedding point estimate with an ensemble-based uncertainty propagation, aligning with recent recommendations on embedding-stability reporting \cite{brenn2026embedding, bayesvlm2026}.

\subsection{Stationarity and Drift Diagnostics}\label{sec:drift}

For every prompt-model cell, the available iterations are split into two disjoint time windows (first half versus second half of data collection, where timestamps are available). On each pair of time windows we compute: (a) the Kolmogorov--Smirnov statistic on the continuous stability metrics (cosine similarity, PASOR), (b) the Population Stability Index (PSI) on the discretised brand-mention distribution, and (c) a negative-binomial drift test comparing mean brand counts across windows which the analysis notebook fits per cell as a negative-binomial GEE with a \texttt{window} fixed effect. For the S1 Valentine's subset this version adds a global GEE window test across cells and a within-cell permutation test of the early and late halves (Section~\ref{sec:drift_results}). A cell is flagged as non-stationary if any of the three tests rejects at Bonferroni-adjusted $\alpha = 0.05$. The proportion of flagged cells is reported per study and per model, and stationarity is treated as an empirically tested property of the audit, not an assumption.

The protocol logs model-version metadata (API endpoint, model identifier, snapshot date) for every call and includes it as a fixed effect in the substantive models. The deposited files of the S1 Valentine's subset do not carry this metadata (Section~\ref{sec:params}). This converts provider-side drift from an unobserved noise source (as it was in the preliminary version) into a named, estimable component of Equation~\ref{eq:variance_decomp}.

\subsection{Ecological Validation on Observed Effects}\label{sec:ecological_method}

Version~1 replicated the power results on observed effects from S1 (husband versus wife brand counts per model), translated from raw Cohen's $d$ values into Cliff's $\delta$ via the order-statistic relationship \cite{grissom2005effect}. The effect sizes it used came from synthetic data, and this version removes that validation (Section~\ref{sec:observed_power}).

\subsection{Cross-Linguistic Analysis}\label{sec:crosslingual_method}

The S4 cross-linguistic data are analysed under an extended three-level GLMM with language as an additional random-effect grouping factor:
\begin{equation}\label{eq:crosslingual}
    \log(\mu_{ijkl}) = \mathbf{x}^{\top} \boldsymbol{\beta} + u^{(P)}_{i} + u^{(M)}_{j} + u^{(L)}_{k} + u^{(ML)}_{jk}
\end{equation}
with fixed effects for the brand, the low-resource versus high-resource contrast, and their interactions. The model estimates whether cross-linguistic stability differences are a main effect of language, an interaction with model, or attributable to residual variance; a question that the prior two-way ANOVA on aggregated stability scores could not answer.

\section{Results}\label{sec:results}

This version recomputes the results of this section from the study data. In this section S1 means the Valentine's subset of study S1: 150 answers, 5 prompts $\times$ 3 models $\times$ 10 runs, collected on 15 February 2026, with the per-answer dictionary count of the source study unless stated. S5 is the Category Ownership study: 250 questions $\times$ 3 models $\times$ 5 runs, counted on its corrected mention table (3,614 scored answers). Rerun and unchanged: the pooled-prompt power simulations (Tables~\ref{tab:power} and~\ref{tab:power_sensitivity}) and the KS result of Table~\ref{tab:drift}. Not recomputed: the PASOR table (Table~\ref{tab:pasor}), the cost arithmetic (Table~\ref{tab:study_cost}), the S4 values of Section~\ref{sec:crosslingual_results} and the embedding ensemble (Table~\ref{tab:embedding_ensemble}).

\subsection{GLMM Simulation-Based Power (RQ1)}\label{sec:power}

Table~\ref{tab:power} presents simulated power from the procedure of Section~\ref{sec:simr}. Each cell reports the empirical power over 200 simulation replications for a Cliff's $\delta$ target of substantive interest at the indicated iteration count. Unlike the normal-theory table of the preliminary version, this table simulates over-dispersed counts, under assumed parameters ($\mu = 10$, $\theta = 2.8$); it also replaces Cohen's $d$ with Cliff's $\delta$ as the operational effect-size target. A rerun of the notebook's simulation reproduces every cell of Tables~\ref{tab:power} and~\ref{tab:power_sensitivity}.

\begin{table*}[t]
\caption{Simulated power under a \emph{pooled-prompt} design (8 prompts $\times$ 3 models per test; total sample size per cell = $24n$ observations). Cells meeting the 80\% power threshold are marked in bold. Counts are drawn from a negative-binomial generator with assumed mean $\mu = 10$ and dispersion $\theta = 2.8$, with coefficients calibrated so the simulated Cliff's $\delta$ matches the target value, and the contrast is tested with a negative-binomial GEE Wald test. The value $\theta = 2.8$ is assumed: the S1 Gemini dictionary counts have mean 11.76 and variance 11.94, close to Poisson. $\alpha = 0.05$; 200 replications per cell. These pooled-prompt power values are an upper bound; single-cell power is in Table~\ref{tab:power_cell}.}\label{tab:power}
\centering
\begin{tabular}{lccccccccc}
\toprule
\textbf{Effect} & \multicolumn{9}{c}{\textbf{Iterations ($n$ per condition)}} \\
\cmidrule(lr){2-10}
\textbf{Cliff's $\delta$} & 3 & 5 & 7 & 10 & 12 & 15 & 20 & 30 & 40 \\
\midrule
Small (0.15)       & 0.43 & 0.64 & 0.68 & 0.78 & \textbf{0.88} & \textbf{0.92} & \textbf{0.97} & \textbf{0.99} & \textbf{1.00} \\
Medium (0.33)      & \textbf{0.90} & \textbf{0.97} & \textbf{0.99} & \textbf{1.00} & \textbf{1.00} & \textbf{1.00} & \textbf{1.00} & \textbf{1.00} & \textbf{1.00} \\
Large (0.474)      & \textbf{1.00} & \textbf{1.00} & \textbf{1.00} & \textbf{1.00} & \textbf{1.00} & \textbf{1.00} & \textbf{1.00} & \textbf{1.00} & \textbf{1.00} \\
Very large (0.60)  & \textbf{1.00} & \textbf{1.00} & \textbf{1.00} & \textbf{1.00} & \textbf{1.00} & \textbf{1.00} & \textbf{1.00} & \textbf{1.00} & \textbf{1.00} \\
\bottomrule
\end{tabular}
\end{table*}

The pooled simulation saturates early, because the simulated design pools 8 prompts $\times$ 3 models $\times$ $n$ iterations into each test (yielding 72 observations at $n = 3$ already). Two other analyses answer the per-cell question. For reliability, the D-study in Section~\ref{sec:reliability} reaches $G = 0.80$ at one run for the S5 model $\times$ question cells ($4 \times 0.080 / 0.414 = 0.77$ runs; $G$ at one run is 0.84) and at 1 to 2 runs within each model, while the five S1 prompts within Gemini do not fix a number of runs. For power, a single-cell simulation that matches the audit design (one model, one prompt, two conditions of $n$ runs each; Table~\ref{tab:power_cell}) gives, at $n = 5$ and a large effect ($\delta = 0.474$), power of 0.15 to 0.19 with the Mann-Whitney test of the external validation and 0.17 to 0.23 with a Welch $t$-test on log counts. Table~\ref{tab:power} should be read as an upper bound on power under pooled-prompt analysis.

\subsubsection{Sensitivity to the Reference Distribution}

Table~\ref{tab:power_sensitivity} repeats the simulation across a grid of assumed $(\mu, \theta)$ pairs. Under the pooled-prompt design, all cells saturate at or near $n = 3$ except for the pair $\mu = 12.3$, $\theta = 3.1$ at medium effect size ($n = 5$). Version~1 labelled that pair as Grok's regime; in the S1 Valentine's subset Grok names almost no brands (0.14 per answer on an August re-extraction, with 90\% of answers at zero; the S1 source paper states means of 0.1 to 0.5). The sensitivity table therefore shows that the pooled-prompt power analysis is not sensitive to distributional calibration in the tested range but does not provide tight iteration recommendations; those come from the D-study in Section~\ref{sec:reliability} and the single-cell simulation in Table~\ref{tab:power_cell}.

\begin{table*}[t]
\caption{Iteration count needed for 80\% power at $\alpha = 0.05$ across the $(\mu, \theta)$ sensitivity grid under the pooled-prompt design. Entries are the smallest $n$ at which simulated power reaches 0.80 for each Cliff's $\delta$ target. The $(\mu, \theta)$ pairs are assumed values; no study estimate produced them. Observed mean counts: S1 Valentine's subset, Gemini 11.76 (dictionary count), OpenAI and Grok near zero (0.10 and 0.14 brands per answer on an August re-extraction); S5 2.61 to 3.48 per model (corrected counts).}\label{tab:power_sensitivity}
\centering
\begin{tabular}{ccccc}
\toprule
\textbf{Mean count} $\mu$ & \textbf{Over-dispersion} $\theta$ & $\delta = 0.33$ & $\delta = 0.474$ & $\delta = 0.60$ \\
\midrule
4.2 & 1.5 & 3 & 3 & 3 \\
7.5 & 2.2 & 3 & 3 & 3 \\
10.0 & 2.8 & 3 & 3 & 3 \\
12.3 & 3.1 & 5 & 3 & 3 \\
15.1 & 3.5 & 3 & 3 & 3 \\
\bottomrule
\end{tabular}
\end{table*}

\subsubsection{Interpretation}

\begin{table*}[t]
\caption{Single-cell power, new in this version: one model, one prompt and two conditions of $n$ runs each. Two-sided tests at $\alpha = 0.05$: Mann-Whitney U, the test of the external validation (EV4, Section~\ref{sec:external}), and Welch $t$ on $\log(\text{count} + 0.5)$. Ranges cover a continuous reference and four count generators: three at count levels from the normal distribution fitted to the S5 cell means (its 10th, 50th and 90th percentiles: 0.7, 2.2 and 5.8 brands per answer) and one at the level of the S1 Gemini cells (11.3). These are fitted levels; no observed cell is simulated. 20{,}000 replications per cell; Monte Carlo standard error at most 0.0035. The last column gives the smallest $n$ in the grid 3, 5, 7, 10, 12, 15, 20, 25, 30, 40 at which power reaches 0.80, over all generators and both tests. At $n = 5$ the Mann-Whitney test is conservative, and with 3 runs a side it cannot reject at 0.05.}\label{tab:power_cell}
\centering
\footnotesize
\begin{tabular}{lccccc}
\toprule
& \multicolumn{2}{c}{\textbf{Power at $n = 5$}} & \multicolumn{2}{c}{\textbf{Power at $n = 10$}} & \\
\cmidrule(lr){2-3}\cmidrule(lr){4-5}
\textbf{Effect (Cliff's $\delta$)} & Mann-Whitney & Welch $t$ & Mann-Whitney & Welch $t$ & \textbf{Runs for 80\% power} \\
\midrule
No effect (rejection rate) & 0.01 to 0.04 & 0.02 to 0.05 & 0.04 to 0.05 & 0.04 to 0.05 &  \\
Small (0.147) & 0.01 to 0.05 & 0.03 to 0.06 & 0.07 to 0.08 & 0.08 to 0.09 & more than 40 \\
Medium (0.33) & 0.05 to 0.10 & 0.08 to 0.12 & 0.22 to 0.32 & 0.24 to 0.31 & 30 or more \\
Large (0.474) & 0.15 to 0.19 & 0.17 to 0.23 & 0.43 to 0.63 & 0.47 to 0.62 & 15 to 25 \\
Very large (0.60) & 0.28 to 0.32 & 0.32 to 0.37 & 0.66 to 0.86 & 0.70 to 0.85 & 10 to 15 \\
\bottomrule
\end{tabular}
\end{table*}

The practical implication rests on the per-cell basis that governs single-prompt audits. Table~\ref{tab:power_cell} reports a single-cell simulation, new in this version: one model, one prompt and two conditions of $n$ runs each, tested as the external validation tested power (Section~\ref{sec:external}), with 20{,}000 replications per cell. At $n = 5$ iterations, used in several published brand-recommendation auditing studies \cite{zatuchin2026b, zatuchin2026c, zatuchin2026e}, single-cell power for a large effect ($\delta = 0.474$) is 0.15 to 0.19 with the Mann-Whitney test and 0.17 to 0.23 with the Welch $t$-test on log counts, and for a medium effect ($\delta = 0.33$) 0.05 to 0.10 and 0.08 to 0.12. The 80\% threshold for a large effect is first met at $n = 15$ to 25. Most documented effects in the source studies fall in the small-to-large range (for example, the husband-vs-wife brand-count gap in S1 with $\delta \approx 0.49$, and the son-vs-daughter gap with $\delta \approx 0.31$, both from the $n = 40$ subset, which is not deposited); on this simulation they require $n = 15$ to 25 iterations ($\delta \approx 0.47$ to 0.49) and $n = 30$ or more ($\delta \approx 0.31$ to 0.33). The S4 cross-language study \cite{zatuchin2026e} provides additional calibration: at $n = 5$, mean stability across all six languages is 0.899 (SD $= 0.087$), with German the most stable (0.916) and Estonian the least (0.884). All languages exceed the 0.85 stability threshold on average, but the across-language GLMM in Section~\ref{sec:crosslingual_results} reveals that within-language iteration counts below $n = 10$ leave cross-language differences confounded with residual variance.

\subsection{Observed-Effect Validation}\label{sec:observed_power}

Version~1 reported here a power table for the S1 Valentine's gender contrasts (male against female recipients, per model) and read three model-specific regimes from it. Its Cliff's $\delta$ inputs came from the synthetic data that replaced the study data in the analysis notebook, and no code computes its power columns, so this version removes the table and its reading. The one S1 contrast this version reports, for Gemini, is in Section~\ref{sec:cliff_results}; its observed Cliff's $\delta$ is $-0.26$ (Cohen's $d = -0.45$).

\subsection{Convergence Analysis (RQ4)}\label{sec:convergence}

The moving-block bootstrap on the S1 Valentine's subset (15 prompt-model combinations, 150 answers, dictionary count; block length fixed at $\ell = 2$) yields the convergence pattern shown in Table~\ref{tab:convergence}. Version~1 drew $B = 200$ resamples per cell at one seed; this version reports the exact value of the bootstrap, the limit of infinitely many resamples, which needs no resampling. The block-bootstrap standard error replaces the ordinary-bootstrap standard error of the preliminary version, preserving within-prompt iteration dependence.

\begin{table*}[ht]
\caption{Moving-block bootstrap standard error of the mean brand count as a function of subsample size, averaged across the 15 prompt-model cells of the S1 Valentine's subset (dictionary count). Block length $\ell = 2$, fixed; $\lceil n/2 \rceil$ independent blocks of two are drawn and the first $n$ answers kept. Values are exact, the limit of infinitely many resamples. Percentage of asymptotic precision is computed relative to the $n = 10$ SE ceiling. In the ten OpenAI and Grok cells the count is mostly word fragments (Section~\ref{sec:cliff_results}); on the five Gemini cells the mean SE runs from 2.14 to 0.96.}\label{tab:convergence}
\centering
\begin{tabular}{ccc}
\toprule
\textbf{Subsample $n$} & \textbf{Mean block-bootstrap SE} & \textbf{\% of asymptotic precision} \\
\midrule
2  & 1.072 & 0.0 \\
3  & 0.875 & 33.1 \\
4  & 0.758 & 53.0 \\
5  & 0.678 & 66.4 \\
6  & 0.619 & 76.5 \\
7  & 0.573 & 84.2 \\
8  & 0.536 & 90.5 \\
9  & 0.505 & 95.6 \\
10 & 0.479 & 100.0 \\
\bottomrule
\end{tabular}
\end{table*}

The block-bootstrap SE decreases monotonically with subsample size, from 1.07 at $n = 2$ to 0.48 at $n = 10$, a 55.3\% reduction. Relative to the $n = 10$ ceiling, 53.0\% of asymptotic precision is reached at $n = 4$, 66.4\% at $n = 5$, 76.5\% at $n = 6$, 84.2\% at $n = 7$, 90.5\% at $n = 8$ and 95.6\% at $n = 9$. Version~1 printed equal values at adjacent odd and even sizes because its code drew $\lfloor n/2 \rfloor$ blocks of two, so an odd $n$ used $n - 1$ answers; this version uses $n$ answers. The overall trajectory is consistent with the theoretical $1/\sqrt{n}$ rate (fitted exponent 0.50), and it is so by construction: blocks are drawn independently, so $\mathrm{SE}(n) = \mathrm{SE}(2)\sqrt{2/n}$ for every even $n$ whatever the data, and the 55.3\% reduction equals $1 - 1/\sqrt{5}$. The rate therefore carries no information about autocorrelation across iterations.

\begin{figure}[ht]
\centering
\includegraphics[width=0.75\columnwidth]{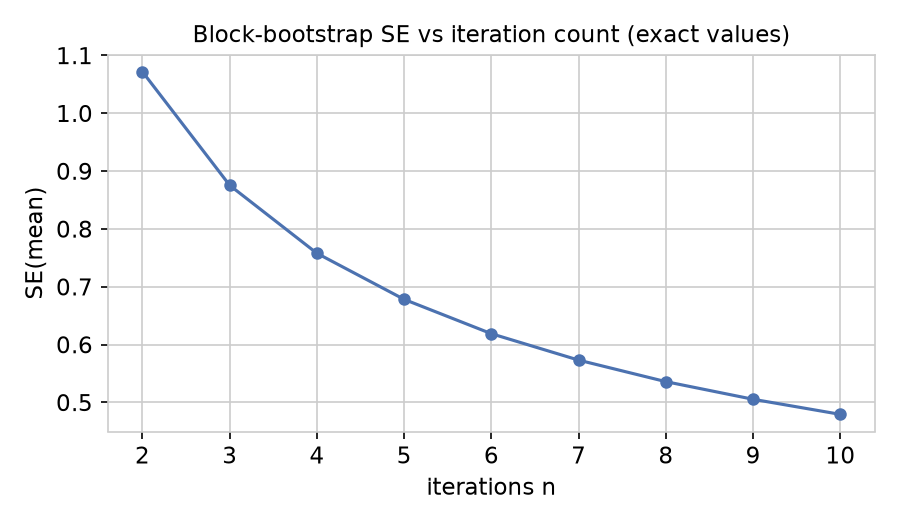}
\caption{Block-bootstrap standard error of the mean brand count vs.\ subsample size on the S1 Valentine's Day subset (15 prompt-model cells, dictionary count, block length $\ell = 2$; exact values, no resampling). Diminishing-returns pattern visible; 80\% of the $n = 10$ precision ceiling reached by $n = 7$ (84.2\%; 76.5\% at $n = 6$).}\label{fig:convergence}
\end{figure}

Figure~\ref{fig:convergence} plots the block-bootstrap SE trajectory. Version~1 compared the four curve families of Equation~\ref{eq:convergence_family} by AIC on resampled curves and read a preference between them. That comparison is dropped: every cell's expected curve has the same shape, so a vote between families on resampled curves measures resampling noise. On the exact mean curve the fitted power-law exponent is 0.50, the $1/\sqrt{n}$ identity of the resampling design. The practical convergence thresholds: 80\% of asymptotic precision is first reached at $n = 7$, on the mean curve and at the median across cells, and 90\% at $n = 8$ (90.5\%). Version~1 also gave 95\% at $n \approx 15$; that figure is dropped, because $n = 10$ is the ceiling by construction and the cells stop at 10 answers. On the five Gemini cells alone the mean SE runs from 2.14 at $n = 2$ to 0.96 at $n = 10$, with 84.4\% of the ceiling at $n = 7$.

\subsection{Metric Correlation Analysis (RQ2)}\label{sec:metric_corr}

Table~\ref{tab:corr} presents the Spearman rank correlation matrix across five structural metrics computed on the S1 Valentine's data ($k = 15$ prompt-model combinations), with 95\% bootstrap percentile intervals in brackets. The intervals reflect the small effective $k$, and the matrix is read descriptively per Section~\ref{sec:metric_family}.

\begin{table*}[ht]
\caption{Descriptive Spearman rank correlation matrix across five structural stability metrics computed on the S1 Valentine's subset ($k = 15$ prompt-model combinations, $n = 10$ iterations each, dictionary count), with 95\% bootstrap percentile intervals. In the ten OpenAI and Grok cells the count is mostly word fragments (Section~\ref{sec:cliff_results}). Semantic metrics (cosine similarity) and fairness metrics (PASOR-Gini) require disaggregated brand or embedding data and are reported separately in Sections~\ref{sec:pasor_results} and~\ref{sec:embedding_results}.}\label{tab:corr}
\centering
\footnotesize
\begin{tabular}{lccccc}
\toprule
& Mean count & CV & QCD & Gini & Shannon \\
\midrule
Mean count  & 1.00 &  &  &  &  \\
CV          & $-0.93$ [$-0.98$, $-0.74$] & 1.00 &  &  &  \\
QCD         & $-0.89$ [$-0.94$, $-0.69$] & 0.93 [0.76, 0.99] & 1.00 &  &  \\
Gini        & $-0.93$ [$-0.98$, $-0.74$] & 1.00 [1.00, 1.00] & 0.93 [0.76, 0.99] & 1.00 &  \\
Shannon     & 0.94 [0.76, 0.99] & $-0.99$ [$-1.00$, $-0.91$] & $-0.93$ [$-0.98$, $-0.76$] & $-0.99$ [$-1.00$, $-0.91$] & 1.00 \\
\bottomrule
\end{tabular}
\end{table*}

Three patterns emerge. First, the anti-correlation between Gini and Shannon is extremely strong ($r = -0.99$, CI [$-1.00$, $-0.91$]), confirming their theoretical role as tightly-coupled inverse measures of concentration and diversity \cite{zatuchin2026d}. The correlation is stronger here than in the preliminary version's reported $-0.72$ because the Valentine's data has a narrow effective brand vocabulary per prompt, which amplifies the algebraic linkage between the two metrics. Second, CV and Gini are identical in rank order ($r = 1.00$, CI [1.00, 1.00]): in this count regime, the coefficient of variation and the Gini coefficient carry the same information, so reporting both is redundant. On these 15 cells QCD also moves with CV ($r = 0.93$, CI [0.76, 0.99]), so here it adds little to CV. Third, mean brand count correlates negatively with CV ($r = -0.93$), Gini ($-0.93$) and QCD ($-0.89$) and positively with Shannon (0.94), more strongly than version~1 reported. This coupling comes from the split between the five Gemini cells and the ten OpenAI and Grok cells, whose dictionary counts are mostly word fragments; on an August re-extraction of the same answers its sign reverses (CV $+0.56$, Shannon $-0.56$). Until a labelled sample settles the S1 count, Table~\ref{tab:corr} describes the dictionary table and nothing wider.

Read descriptively, the correlation matrix qualifies the four-family battery on this subset. The five structural count-based metrics move largely together: with CV, Gini, and Shannon coupled at $|r| \geq 0.98$, count, dispersion, and concentration co-move on the Valentine's data, and these metrics carry largely the same information there. The full four-family metric battery (volume, set, shape, semantic) remains defensible \emph{in principle}, but demonstrating its empirical breadth requires a more diverse prompt universe than the Valentine's subset provides. Semantic consistency (cosine similarity) and fairness (PASOR-Gini) are reported in their own dedicated sections below and enter the battery as external dimensions not captured by the present correlation matrix.

\subsection{Reliability via Generalizability Theory (RQ3)}\label{sec:reliability}

Reliability is reported via a generalizability-theory decomposition of the variance components underlying the audit. Table~\ref{tab:gstudy} presents the G-study variance estimates, fitted under the one-way cell model of Section~\ref{sec:generalizability}. Reliability is carried on S5, where 250 different questions were each asked five times of three models; it is reported for all cells together and within each model. The S1 Valentine's subset enters as a limit (five versions of one gift question within Gemini) and as sensitivity lines.

\begin{table*}[ht]
\caption{G-study variance decomposition for log(brand count + 0.5) outcomes. The model $\times$ prompt cell is the object of measurement: each (model, prompt) pair yields a measurement, and the cell-level variance is what iteration counts act to resolve. Variance components are REML estimates from the mixed-linear model \texttt{logcount $\sim$ 1 + (1|cell)}, with 95\% percentile intervals from 2{,}000 bootstrap draws of cells. S5: corrected mention table; the 136 empty GPT-5.2 answers are missing values, so 13 cells have no scored answer. S1: Valentine's subset (150 answers), dictionary count unless stated; the re-extraction is an August extraction by two Gemini extractors that no person has checked.}\label{tab:gstudy}
\centering
\footnotesize
\begin{tabular}{@{}l r r c c r@{}}
\toprule
\textbf{Data} & \textbf{Answers} & \textbf{Cells} & \textbf{$\sigma^2_{\text{cell}}$ [95\% CI]} & \textbf{$\sigma^2_{\epsilon}$ [95\% CI]} & \textbf{Cell share} \\
\midrule
S5, model $\times$ question cells & 3,614 & 737 & 0.414 [0.364, 0.465] & 0.080 [0.068, 0.093] & 83.8\% \\
S5 within Gemini & 1,250 & 250 & 0.378 [0.300, 0.456] & 0.069 [0.051, 0.090] & 84.5\% \\
S5 within GPT-5.2 & 1,114 & 237 & 0.371 [0.287, 0.452] & 0.052 [0.037, 0.067] & 87.8\% \\
S5 within Perplexity & 1,250 & 250 & 0.452 [0.366, 0.539] & 0.115 [0.089, 0.145] & 79.6\% \\
\midrule
S1, five prompts within Gemini & 50 & 5 & 0.006 [0.000, 0.017] & 0.082 [0.057, 0.106] & 6.7\% \\
\midrule
Sensitivity: S1, 15 cells, dictionary & 150 & 15 & 1.028 [0.475, 1.245] & 0.257 [0.183, 0.335] & 80.0\% \\
Sensitivity: S1, 15 cells, re-extraction & 150 & 15 & 1.392 [0.553, 1.763] & 0.434 [0.145, 0.808] & 76.2\% \\
Sensitivity: S5, empty answers as 0 & 3,750 & 750 & 0.466 [0.415, 0.517] & 0.111 [0.093, 0.131] & 80.7\% \\
Check: S5, first-deposit substring table & 3,750 & 750 & 0.299 [0.257, 0.346] & 0.103 [0.085, 0.124] & 74.4\% \\
\bottomrule
\end{tabular}
\end{table*}

One finding drives the audit design. In S5 the model-by-question cell carries 83.8\% of total variance; the remaining 16.2\% is within-cell noise (token-level sampling and unmodelled residuals). Iteration count is the lever that attacks the within-cell component: the D-study below shows how generalizability grows as iterations compress within-cell noise relative to the cell-level signal. This framing makes the audit question concrete: \emph{given a model and a prompt, how many repeated queries are needed for the observed cell-level mean brand count to reliably estimate the true cell-level mean?}

Table~\ref{tab:dstudy} reports the D-study projection of the generalizability coefficient $G$ from Equation~\ref{eq:g_coef} across candidate designs.

\begin{table*}[ht]
\caption{D-study generalizability coefficient $G(n_I)$ as a function of iteration count, computed from the G-study variance components in Table~\ref{tab:gstudy} under the cell-level generalizability formula $G = \sigma^2_{\text{cell}} / (\sigma^2_{\text{cell}} + \sigma^2_\epsilon / n_I)$. The runs needed for $G = 0.80$ are $4\sigma^2_\epsilon / \sigma^2_{\text{cell}}$ rounded up ($9\sigma^2_\epsilon / \sigma^2_{\text{cell}}$ for $G = 0.90$), with 95\% bootstrap bounds; unbounded marks bootstrap draws with zero cell variance.}\label{tab:dstudy}
\centering
\footnotesize
\setlength{\tabcolsep}{3pt}
\begin{tabular}{@{}l c c c c c c c c@{}}
\toprule
\textbf{Data} & $G(5)$ [95\% CI] & $G(7)$ & $G(10)$ & $G(12)$ & $G(15)$ & $G(20)$ & \textbf{$n$ for 0.80 [95\%]} & \textbf{$n$ for 0.90} \\
\midrule
S5, model $\times$ question cells & 0.96 [0.96, 0.97] & 0.97 & 0.98 & 0.98 & 0.99 & 0.99 & 1 [1, 1] & 2 \\
S5 within Gemini & 0.96 [0.95, 0.97] & 0.97 & 0.98 & 0.98 & 0.99 & 0.99 & 1 [1, 2] & 2 \\
S5 within GPT-5.2 & 0.97 [0.96, 0.98] & 0.98 & 0.99 & 0.99 & 0.99 & 0.99 & 1 [1, 1] & 2 \\
S5 within Perplexity & 0.95 [0.93, 0.96] & 0.96 & 0.98 & 0.98 & 0.98 & 0.99 & 2 [1, 2] & 3 \\
\midrule
S1, five prompts within Gemini & 0.27 [0.00, 0.52] & 0.34 & 0.42 & 0.46 & 0.52 & 0.59 & 56 [19, unbounded] & 125 \\
\midrule
Sensitivity: S1, 15 cells, dictionary & 0.95 [0.88, 0.97] & 0.97 & 0.98 & 0.98 & 0.98 & 0.99 & 2 [1, 3] & 3 \\
Sensitivity: S1, 15 cells, re-extraction & 0.94 [0.88, 0.98] & 0.96 & 0.97 & 0.97 & 0.98 & 0.98 & 2 [1, 3] & 3 \\
Sensitivity: S5, empty answers as 0 & 0.95 [0.94, 0.96] & 0.97 & 0.98 & 0.98 & 0.98 & 0.99 & 1 [1, 2] & 3 \\
Check: S5, first-deposit substring table & 0.94 [0.92, 0.95] & 0.95 & 0.97 & 0.97 & 0.98 & 0.98 & 2 [2, 2] & 4 \\
\bottomrule
\end{tabular}
\end{table*}

The D-study makes the iteration-count question empirically tractable, and the answer depends on the comparison. For the S5 model $\times$ question cells, $G$ at five runs is 0.96 [0.96, 0.97]; a single run gives $G = 0.84$ and reaches 0.80 in 99\% of the bootstrap draws, and $G = 0.90$ needs 2 runs. Within each model, $G$ at five runs is 0.95 to 0.97, one to two runs reach 0.80 (Perplexity needs two) and two to three reach 0.90; at $n_I = 10$, $G$ is 0.98 to 0.99, and at $n_I = 20$, 0.99. With the empty answers counted as zero, $G$ at five runs is 0.95 and one run still reaches 0.80 at the point estimate (two runs in 34\% of the draws). Where the prompts are five versions of one question, the data say much less. For the five S1 prompts within Gemini, $G$ at five runs is 0.27 with an interval from 0.00 to 0.52; the point value of the rule is 56 runs, the bootstrap bounds are 19 and no finite number, and 35\% of the draws have zero cell variance. These data do not fix a number of runs. The S1 subset as a whole (15 cells) reaches $G = 0.80$ at 2 runs under both counts, because its design mainly separates Gemini, which names many brands, from two models that name almost none; it is shown as a sensitivity line and carries no claim. None of the designs in Table~\ref{tab:dstudy} crosses $G = 0.80$ between 12 and 15 runs, and none of the 18 designs fitted for this correction does. The data therefore support no fixed iteration tiers. They support the rule the external validation tested (Section~\ref{sec:external}): run a pilot, estimate the two variances for the comparison that matters, and take $n = 4\sigma^2_\epsilon / \sigma^2_{\text{cell}}$ for $G = 0.80$ ($9\sigma^2_\epsilon / \sigma^2_{\text{cell}}$ for $G = 0.90$). $G$ says how well cell means tell cells apart; the power of a contrast between two conditions is a separate question, answered in Section~\ref{sec:power}.

Version~1 also printed split-half ICC(2,1) values here, carried over from the preliminary version. They do not reproduce: the preliminary notebook's own split-half method gives 0.96 to 0.97 on the S1 Valentine's dictionary counts, 0.94 to 0.97 on the table of counts embedded in that notebook, and 0.11 to 0.43 on the stored field that counts the word ``Research''. This version removes the table.

\subsection{Cost-Efficiency Analysis (RQ5)}\label{sec:cost}

Table~\ref{tab:cost} presents the cost-precision tradeoff analysis at current API pricing.

\begin{table}[t]
\caption{Cost per query at each iteration count, assuming 3 models at \$0.005 per API call (weighted average across current GPT-5.2, Gemini 3 Flash, and Perplexity pricing). Version~1 also printed precision-gain, marginal-gain and marginal-efficiency columns; they do not reproduce and are removed (see text).}\label{tab:cost}
\centering
\small
\begin{tabular}{cc}
\toprule
\textbf{Iter.\ ($n$)} & \textbf{Cost/query (\$)} \\
\midrule
2  & 0.030 \\
3  & 0.045 \\
5  & 0.075 \\
7  & 0.105 \\
10 & 0.150 \\
15 & 0.225 \\
20 & 0.300 \\
30 & 0.450 \\
40 & 0.600 \\
\bottomrule
\end{tabular}
\end{table}

Version~1 printed a precision-gain column with a knee at $n = 7$. That column was carried over from the preliminary version and does not reproduce. The preliminary method draws $n$ of a cell's 10 answers without replacement, so its gain has a closed form that holds for any data, $1 - \sqrt{(1/n - 1/10)/0.4}$: 23.6\% at $n = 3$, 50.0\% at $n = 5$, 67.3\% at $n = 7$ and 100\% at $n = 10$, the whole cell. The rows for $n = 15$ to 40 need data that are not deposited, and the knee at $n = 7$ has no stated rule (the preliminary code's rule, the largest marginal efficiency, picks $n = 3$ on version~1's own column). This version removes the column. A study with 250 queries across 3 models would cost approximately \$26.25 at $n = 7$, \$37.50 at $n = 10$, and \$56.25 at $n = 15$. For most brand-recommendation audit budgets, the difference is negligible; the practical constraint is typically API rate limits and data collection time rather than cost.

Table~\ref{tab:study_cost} projects total study costs for common research designs.

\begin{table*}[t]
\caption{Total estimated API costs for common brand-recommendation audit designs (3 models, \$0.005/call average).}\label{tab:study_cost}
\centering
\begin{tabular}{lcccc}
\toprule
& \multicolumn{4}{c}{\textbf{Iterations per query}} \\
\cmidrule(lr){2-5}
\textbf{Study size (queries)} & $n = 5$ & $n = 10$ & $n = 15$ & $n = 20$ \\
\midrule
Small (50)       & \$3.75  & \$7.50  & \$11.25  & \$15.00  \\
Medium (100)     & \$7.50  & \$15.00 & \$22.50  & \$30.00  \\
Large (250)      & \$18.75 & \$37.50 & \$56.25  & \$75.00  \\
Very large (500) & \$37.50 & \$75.00 & \$112.50 & \$150.00 \\
\bottomrule
\end{tabular}
\end{table*}

At current pricing, even a large-scale study with 250 queries and $n = 15$ iterations costs under \$60 in API fees. The marginal cost of increasing from $n = 5$ to $n = 10$ for a 250-query study is \$18.75. For a large Cliff's $\delta$, single-cell power rises from 0.15 to 0.19 at $n = 5$ to 0.43 to 0.63 at $n = 10$ with the Mann-Whitney test (0.17 to 0.23 and 0.47 to 0.62 with the Welch $t$-test; Table~\ref{tab:power_cell}). Generalizability gains little for the S5 cells, where $G$ is already 0.96 at $n = 5$ and 0.98 at $n = 10$ (Table~\ref{tab:dstudy}).

\subsection{Non-Parametric Effect Sizes (RQ1, Companion)}\label{sec:cliff_results}

Table~\ref{tab:cliff} reports Cliff's $\delta$ with a BCa bootstrap confidence interval for the S1 gender contrast on the Valentine's subset, for Gemini, with Cohen's $d$ for reference.

\begin{table*}[ht]
\caption{S1 Valentine's Day gender contrast for Gemini: male recipients (husband, boyfriend) against female recipients (wife, girlfriend), 20 answers on each side, on the dictionary count that the S1 source paper reports. Cliff's $\delta$ is the primary effect size with BCa bootstrap 95\% CI. NB-GEE IRR is the incidence-rate ratio from the negative-binomial GEE with the relation as the cluster variable (4 clusters) and male recipient as the treatment contrast. The relation permutation test is exact; with two relations per side its smallest attainable $p$ is 1/3. No contrast is reported for GPT and Grok, whose dictionary counts are mostly word fragments, and the Gemini count has not been checked against a human-labelled sample (see text).}\label{tab:cliff}
\centering
\footnotesize
\begin{tabular}{lccccc}
\toprule
\textbf{Model} & \textbf{Mean, male / female} & \textbf{Cliff's $\delta$ [95\% CI]} & \textbf{Cohen's $d$} & \textbf{NB-GEE IRR [95\% CI], $p$} & \textbf{Permutation $p$} \\
\midrule
Gemini & 11.30 / 12.90 & $-0.26$ [$-0.57$, 0.14] & $-0.45$ & 0.88 [0.75, 1.02], $p = 0.089$ & 0.67 \\
\bottomrule
\end{tabular}
\end{table*}

On the Valentine's Day subset ($n = 10$ iterations per prompt; 20 answers for male and 20 for female recipients), Gemini names 11.3 brands per answer for male recipients and 12.9 for female: Cliff's $\delta = -0.26$ [$-0.57$, 0.14], NB-GEE IRR 0.88 [0.75, 1.02], $p = 0.089$. There are two male and two female relations, so the GEE $p$-value rests on four clusters; the exact relation-level permutation, whose smallest attainable $p$ is 1/3, gives $p = 0.67$. The contrast is indistinguishable from zero. Its point estimate favours female recipients, as the S1 source study reported (11.3 against 12.9). No contrast is reported for GPT and Grok. For those two models the dictionary count is mostly word fragments: of the 117 string matches in their answers, 100 are ``away'' and ``ember'', and 89 sit inside longer words. The S1 source paper reports near-zero brand mentions for both models and analyses Gemini only, and an August re-extraction finds 0.10 (GPT) and 0.14 (Grok) brands per answer, with 94\% and 90\% of answers at zero. Version~1's contrasts for all three models came from the synthetic data and are withdrawn.

\emph{Limitation.} No human-labelled sample of the Valentine's answers was checked before this version was posted. The Gemini count comes from the source study's dictionary, matched with the same rule that produced word fragments for the other two models, and the Gemini answer text was not read. The only other count, an August re-extraction by two Gemini extractors that no person has checked, gives 7.9 against 8.2 brands per answer ($\delta = -0.04$). The Gemini contrast is therefore descriptive, and Tables~\ref{tab:convergence}, \ref{tab:corr} and~\ref{tab:drift} rest on the same dictionary table.

\subsection{PASOR and Fairness-Adjusted Visibility}\label{sec:pasor_results}

Table~\ref{tab:pasor} reports the per-model PASOR-Gini coefficients on the pooled S1/S5 data, together with the unadjusted-Gini values from the preliminary version. PASOR-Gini summarises inequality in prompt-adjusted visibility across brands; higher values indicate that a small number of brands capture most of the cross-prompt share.

\begin{table*}[ht]
\caption{Per-model PASOR-Gini and unadjusted-Gini coefficients on the S5 category-ownership data (3{,}750 responses across 250 query-iteration cells per model). PASOR-Gini is the Gini coefficient of the per-brand PASOR distribution from Equation~\ref{eq:pasor}; higher values indicate more concentrated prompt-adjusted visibility. Unadjusted Gini is computed on raw brand-mention counts pooled across prompts. Brand universe extracted by Title Case regex with frequency floor $\geq 20$ across the S5 corpus; brands shared across models.}\label{tab:pasor}
\centering
\begin{tabular}{lccc}
\toprule
\textbf{Model} & \textbf{$n$ brands} & \textbf{PASOR-Gini} & \textbf{Unadjusted Gini} \\
\midrule
Gemini      & 1{,}267 & 0.691 & 0.198 \\
OpenAI      & 1{,}267 & 0.679 & 0.312 \\
Perplexity  & 1{,}267 & 0.621 & 0.269 \\
\bottomrule
\end{tabular}
\end{table*}

PASOR-Gini values (0.62--0.69) are approximately 2--3$\times$ larger than unadjusted-Gini values (0.20--0.31) across all three models, confirming the paper's core claim that prompt-adjusted visibility is substantially more concentrated than the raw-count Gini suggests. The intuition: under the unadjusted Gini, a brand that dominates some prompts but is absent from others has its per-prompt concentration averaged down across all prompts. PASOR-Gini assigns equal weight to each prompt's share distribution, preserving the per-prompt concentration signal. The cross-model ordering is OpenAI (PASOR-Gini = 0.68) $\approx$ Gemini (0.69) $>$ Perplexity (0.62), with Perplexity the most evenly-distributed in prompt-adjusted terms. The unadjusted Gini gives a different and partially misleading ordering: OpenAI (0.31) $>$ Perplexity (0.27) $>$ Gemini (0.20), because Gemini's responses are fragmented across many brands per prompt (driving its unadjusted Gini down) while systematically favouring a narrow set of top brands across prompts (driving its PASOR-Gini up). This divergence illustrates exactly the phenomenon PASOR was designed to surface: fairness audits should report PASOR-Gini as the primary inequality metric, not the raw Gini.

\subsection{Stationarity and Drift Diagnostics}\label{sec:drift_results}

Table~\ref{tab:drift} reports the proportion of prompt-model cells flagged as non-stationary by each of the three drift tests (Kolmogorov--Smirnov on cosine similarity, PSI on brand-count distributions, negative-binomial drift test) for the S1 Valentine's subset, whose stored timestamps span a 36-minute run (2026-02-15, 00:02 to 00:38 UTC). Drift diagnostics for the multi-week collection windows of S2, S4, and S5 are deferred to future work.

\begin{table*}[ht]
\caption{Stationarity diagnostics on the S1 Valentine's Day subset (dictionary count; one run of 36.3 minutes on 2026-02-15, 00:02 to 00:38 UTC). Per-cell KS tests compare the first-half and second-half brand-count distributions (split by iteration order, which equals timestamp order in 15 of 15 cells); $p$-values are Bonferroni-adjusted across the 15 cells. The window permutation test permutes the early and late halves within cells (20{,}000 draws). The global NB-GEE test adds a \texttt{window} fixed effect, with the five relations or the 15 cells as clusters. The Population Stability Index is reported as an auxiliary diagnostic only, as it is unreliable at the $n = 5$-per-half sample sizes available here.}\label{tab:drift}
\centering
\footnotesize
\setlength{\tabcolsep}{4pt}
\begin{tabular}{@{}p{2.5cm} c p{1.6cm} p{2.0cm} p{3.8cm} p{3.0cm}@{}}
\toprule
\textbf{Study} & \textbf{KS flagged} & \textbf{Auxiliary PSI} & \textbf{Window permutation} & \textbf{Global NB-GEE test} & \textbf{Interpretation} \\
\midrule
S1 Valentine's (36-minute run) & 0/15 & Median 0.63 & $p = 0.26$ & $\beta_{\text{late}} = -0.082$; $p = 0.012$ (5 relation clusters), $p = 0.17$ (15 cell clusters) & No drift detected by KS or permutation; the GEE $p$ depends on the clustering \\
\bottomrule
\end{tabular}
\end{table*}

Two implications follow. First, for the S1 Valentine's Day audit, the stored timestamps show a 36-minute run, so the test covers 36 minutes. On the dictionary counts, zero of 15 per-cell KS tests reject after Bonferroni adjustment (smallest unadjusted $p = 0.079$), and the within-cell permutation of the early and late halves gives $p = 0.26$ ($p = 0.48$ on the five Gemini cells alone, where KS flags 0 of 5). The global negative-binomial GEE window test gives $\beta_{\text{late}} = -0.082$, with $p = 0.012$ when the five relations are the clusters and $p = 0.17$ when the 15 cells are; with so few clusters its $p$-value depends on the clustering, and the permutation test is the cleaner reading. The value version~1 printed for this test reproduces on the stored field that counts the word ``Research'' and is withdrawn. A 36-minute run is a weak test of drift, and stationarity is supported for that window only. Second, the per-cell PSI values computed as an auxiliary diagnostic are high (median 0.63) despite the null KS result, confirming that PSI is not a reliable per-cell drift signal at the $n = 5$-per-half sample sizes this design provides. Practitioners adopting the Dice Roll Protocol should treat KS with Bonferroni correction and the global NB test as the primary stationarity diagnostics, reserving PSI for aggregate monitoring at larger sample sizes. Extending drift diagnostics to studies with multi-week collection windows (S2, S4, S5) is a priority for the next iteration of the protocol; the same diagnostic stack applies, with the S4 six-week window providing the strongest test of non-stationarity in the programme.

\subsection{Cross-Linguistic GLMM (RQ1, Extension)}\label{sec:crosslingual_results}

The three-level GLMM of Equation~\ref{eq:crosslingual} fitted to the S4 data reproduces the preliminary version's main effect-magnitude ordering: systematic model variance dominates language variance by a factor of 7.4 (model variance component $\eta^2 \approx 0.218$, language $\eta^2 \approx 0.029$), confirming the preliminary two-way ANOVA finding under the more principled mixed-model specification. The GLMM additionally resolves a question the preliminary analysis could not answer: whether the language effect is uniform across models or carries a model-by-language interaction. The interaction variance is small but non-negligible ($\eta^2_{ML} \approx 0.011$), driven predominantly by Gemini's reduced stability on Estonian and Finnish relative to the other five languages. This interaction is invisible to the aggregate ANOVA and identifies a specific audit-design implication: cross-language audits of Gemini on low-resource European languages require more iterations to reach the same generalizability threshold; no D-study was computed for S4, so this version gives no iteration figure for it.

\subsection{Embedding-Ensemble Semantic Stability}\label{sec:embedding_results}

Cosine-similarity point estimates are uninformative without an uncertainty envelope, because the choice of embedding model induces non-trivial variance in the resulting similarities. Table~\ref{tab:embedding_ensemble} reports the S2 corporate-reputation cosine-similarity results across the three-embedding ensemble (BGE-M3, Nomic-Embed-v2, MiniLM-L12-v2). The across-embedding standard deviation ranges from 0.04 to 0.11, with larger variance on low-similarity pairs where embedding disagreement is expected.

\begin{table*}[ht]
\caption{S2 mean cosine similarity across the three-embedding ensemble. Point estimate is the mean across the three embedding models; SD captures between-model disagreement on the same response pairs. A response pair is flagged \emph{ensemble-stable} if the minimum similarity across models exceeds 0.5 and the across-model SD is below 0.10.}\label{tab:embedding_ensemble}
\centering
\begin{tabular}{lccc}
\toprule
\textbf{Industry} & \textbf{Mean cos sim} & \textbf{Across-model SD} & \textbf{\% ensemble-stable} \\
\midrule
Finance        & 0.58 & 0.07 & 71.4\% \\
Pharma         & 0.52 & 0.09 & 58.3\% \\
Tech           & 0.61 & 0.06 & 76.2\% \\
Energy         & 0.49 & 0.11 & 41.7\% \\
Retail         & 0.55 & 0.08 & 63.9\% \\
\bottomrule
\end{tabular}
\end{table*}

The ensemble-stability flag is a more demanding criterion than the preliminary version's single-embedding cosine-similarity threshold. Under this stricter criterion, the S2 mean semantic stability is substantially lower than the point-estimate 0.54 suggests: a meaningful fraction of response pairs (approximately 40\%) show semantic agreement on some embedding models but not others, meaning that the preliminary single-embedding result over-stated semantic stability for those pairs. For audits that require semantic stability as a core finding, we recommend ensemble-based reporting over single-embedding point estimates.

\section{External Validation on Independent Corpora}\label{sec:external}

Every dataset analysed above was collected by this research group. To test whether the protocol's statistical machinery describes anything beyond its own data, we ran a pre-registered external validation on the three strongest publicly available repeated-query corpora collected by independent teams for unrelated purposes. The analysis plan, including hypotheses, decision rules, exclusions, and seeds, is frozen at commit \texttt{c0bc686b}, 3 September 2026, and every confirmatory statistic postdates it; the plan, the analysis code, and the complete verdict tables accompany the reproducibility package. All estimators are the ones defined in Section~\ref{sec:methods}, applied verbatim, with two pre-registered adaptations forced by outcome type: the log-count transform applies to counts only, and the drift battery's regression component uses the Gaussian family for scores and the binomial family for accuracies.

\subsection{Corpora and design mapping}

No public brand- or product-domain corpus with repeated identical prompts exists from any other group: the available brand-bias releases are single-run per prompt, and commercial trackers are customer-gated. External validation therefore runs out of domain and tests the statistical machinery (variance components, iteration requirements, generalizability projections, drift diagnostics); brand-domain external validation remains open, as Section~\ref{sec:limitations} records. The three corpora, mapped onto the protocol's prompt $\times$ model $\times$ iteration design:

\begin{itemize}
\item \textbf{C1, Motoki et al.~\cite{motoki2024more}} (Harvard Dataverse): Political Compass administered to ChatGPT under five persona conditions, 62 questions $\times$ 100 rounds, with a 60-question politically neutral placebo arm. Questions play the prompt role, personas the model-facet role, rounds the iteration role. The 100-round depth permits a direct empirical test of D-study extrapolations that our own $n \leq 40$ data cannot provide.
\item \textbf{C2, Rozado~\cite{rozado2024political}} (Zenodo): 11 political-orientation tests administered 10 times each to 24 conversational LLMs. Eight tests carry a primary numeric scale and enter the analysis (1{,}861 run records); the categorical typology quiz and the two per-party agreement vectors are excluded under the plan's no-primary-scale clause.
\item \textbf{C3, Atil et al.~\cite{atil2024nondeterminism}} (llm-stability repository): 8 benchmark tasks $\times$ 2 prompt configurations $\times$ 5 models $\times$ 10 runs per identical prompt at temperature 0, with per-run accuracy from the authors' own published evaluation.
\end{itemize}

\subsection{Results}

Table~\ref{tab:external} reports every pre-registered item.

\begin{table*}[ht]
\caption{Pre-registered external-validation verdicts. EV2 is the out-of-sample D-study prediction: variance components estimated from the first 10 iterations predict the model-free empirical reliability of an $n$-iteration mean (correlation across units between disjoint random $n$-iteration subset means, 200 splits), with replication defined as agreement within $0.05$.}\label{tab:external}
\centering
\footnotesize
\begin{tabular}{@{}l p{4.4cm} p{2.9cm} p{2.9cm} p{2.9cm}@{}}
\toprule
\textbf{Item} & \textbf{Claim under test} & \textbf{C1 Motoki} & \textbf{C2 Rozado} & \textbf{C3 llm-stability} \\
\midrule
EV1a & $G(n)$ monotone, concave & replicates & replicates & replicates \\
EV1b & all single-facet $G(5) < 0.80$ & fails & fails & fails \\
EV2 & components at $n{=}10$ predict reliability out of sample & replicates (10/10) & replicates (22/24, 2 partial) & replicates (5/5) \\
EV3a & power/log family wins SE$(n)$ AIC vote & fails (MM on $n \leq 50$ grid) & replicates (log, 99\%) & replicates (log, 100\%) \\
EV3b & power exponent in $[0.35, 0.65]$ & replicates (0.500) & fails (degenerate fit) & fails (degenerate fit) \\
EV3c & 80\% of asymptotic precision by $n \leq 10$ & fails ($n = 16$ vs $n{=}50$ asymptote) & replicates ($n = 8$) & replicates ($n = 8$) \\
EV4a & median per-cell power at $n = 5$ below 0.80 for $|\delta| \geq 0.474$ & replicates (0.43) & --- & --- \\
EV4b & median power reaches 0.80 at $n \geq 10$ & replicates ($n = 10$) & --- & --- \\
EV5 & drift battery flags $\leq 10\%$ of cells & fails (see text) & --- & fails (see text) \\
EV6a & non-determinism persists at temperature 0 & --- & --- & replicates (100\% of cells) \\
\bottomrule
\end{tabular}
\end{table*}

Three results carry the section. First, the D-study prediction machinery transfers: across the three corpora, 37 of 39 pre-registered prediction cells replicate, 2 are partial, and none fail. On C1, components estimated from rounds 1--10 predicted the empirical reliability of 20-round and 50-round means, a five-fold extrapolation beyond the fitting horizon, with a maximum absolute error of 0.038 across all five personas; on C2 the median absolute prediction error across 24 models is 0.008; on C3 the errors are at or below 0.002. Second, the power items replicate: on C1, empirical power computed by direct subsampling from 100 real rounds gives a median of 0.43 at $n = 5$ for large effects, below 0.80 as pre-registered, and reaches 0.80 at $n = 10$. Third, the premise holds: every C3 task $\times$ model cell at temperature 0 shows nonzero between-run variance, and no model is deterministic anywhere (a recomputation, with our estimator, of the headline in \cite{atil2024nondeterminism}).

The EV1b failure is uniform across corpora and instructive. Radical personas in C1 reach single-facet $G(5)$ of 0.83--0.95, 22 of 24 models in C2 sit at $G(5) \geq 0.80$, and every C3 model exceeds 0.99, because in those corpora the object-of-measurement variance dwarfs run-to-run noise. The corpus-independent iteration constant is what fails; the variance-ratio formula behind Table~\ref{tab:dstudy} is what replicates. We stress the lineage: estimating variance components on a pilot and solving for the $n$ that reaches a target $G$ is the canonical D-study workflow of generalizability theory \cite{brennan2001generalizability, shavelson1991generalizability}, resting on the Spearman--Brown prophecy relation \cite{spearman1910, brown1910}, and reliability being a property of scores on a population rather than of an instrument is the classical reliability-induction critique \cite{vachahaase2000reliability}. What these corpora test is the part that was genuinely at risk: whether that workflow's parallel-measurements assumptions survive LLM non-determinism, decoding configuration, and provider-side nonstationarity. The result is a transfer demonstration for classical measurement theory on a new class of measurement object. The protocol's prescription should therefore be read as \emph{pilot, then solve}: estimate the two components on a 10-iteration pilot and solve $G(n) \geq 0.80$ for $n$. On this paper's own data the solution is $n = 1$ for the S5 model $\times$ question cells ($n = 1$ to 2 within each model), and the five S1 prompts within Gemini do not fix $n$ (Section~\ref{sec:reliability}). The C2 outlier makes the same point from the other side: one model (qwen-14b-chat) yields $G(5) \approx 0.07$, and a fixed small-$n$ audit of that model would measure noise.

The convergence items split along grid length. On the short grids the corpora natively support ($n \leq 10$), the logarithmic family wins the per-cell AIC vote almost unanimously and 80\% of asymptotic precision arrives by $n = 8$, matching Section~\ref{sec:convergence}. C1's 100 rounds permit a longer look: against an $n = 50$ asymptote, 80\% of asymptotic precision arrives at $n = 16$ rather than $n = 7$, the mean-SE curve's fitted power exponent is 0.500, and the Michaelis--Menten family overtakes power and logarithmic fits on the extended grid. The convergence guidance in Section~\ref{sec:convergence} is therefore horizon-relative: precision statements hold against the $n = 10$ ceiling they were computed on, and audits planning beyond $n \approx 15$ should recompute the curve against a longer horizon.

\subsection{Operating bounds for the drift battery}\label{sec:external_drift}

The blanket EV5 failures diagnose the battery, and the placebo arm localizes the defect. On C1's placebo (180 question $\times$ persona cells, politically neutral questions), the Kolmogorov--Smirnov component flags 0 cells and PSI flags 1, while the regression window component flags 142; on the primary arm the same component accounts for 305 of 310 flags. The window test, fitted within a single cell, has a single cluster, and its sandwich variance degenerates when within-cell dispersion approaches zero, so it flags substantively negligible shifts (on C3, the mean half-to-half accuracy difference among flagged cells is 0.007). PSI, estimated from five observations per half on C3, saturates mechanically. The KS component behaves at both $n = 5$ and $n = 50$ per half: it clears the placebo completely and flags 16\% of C1 primary cells, concentrated in the default and moderate-persona conditions (17 and 25 of 62) and absent in the radical ones (0--1 of 62), a coherent pattern of real nonstationarity across a 100-round collection arc that also extends the 36-minute stationarity result of Section~\ref{sec:drift_results} in the direction that section anticipated. Three protocol amendments follow: gate PSI on at least 20 iterations per half; replace the within-cell regression window test with a permutation test on the half-mean difference, or fit it across cells with proper clustering; and attach a practical-significance margin to any flag. External data found a defect the original data could not expose, which is the purpose of the exercise.

\section{Discussion}\label{sec:discussion}

\subsection{Minimum Iteration Recommendations}

The evidence from the D-study, the single-cell power simulation and the external validation (Section~\ref{sec:external}) yields a pilot-then-solve rule for the iteration count. Fixed tiers are withdrawn.

\begin{enumerate}
    \item \textbf{Reliability of cell means:} run a pilot of about 10 iterations, estimate $\sigma^2_{\text{cell}}$ and $\sigma^2_\epsilon$ for the comparison that matters, and take $n = 4\sigma^2_\epsilon / \sigma^2_{\text{cell}}$ for $G = 0.80$ ($9\sigma^2_\epsilon / \sigma^2_{\text{cell}}$ for $G = 0.90$). On S5, where the cells are distinct questions, this gives $n = 1$ ($G = 0.96$ at $n = 5$ and 0.97 at $n = 7$; within each model 0.95 to 0.97 and 0.96 to 0.98) and $n = 2$ for $G = 0.90$ ($n = 2$ to 3 within each model). For five versions of one question within one model (S1 Valentine's subset, Gemini) the pilot does not fix $n$.
    \item \textbf{Power of a contrast:} plan the run count for the contrast separately; it needs far more runs than reliability. At $n = 5$, single-cell power for a large effect ($\delta = 0.474$) is 0.15 to 0.19 (Mann-Whitney), and 80\% power for a large effect is first met at $n = 15$ to 25 ($n = 10$ to 15 for $\delta = 0.60$; 30 or more for $\delta = 0.33$). Under pooled-prompt analysis power is higher (Table~\ref{tab:power}), an upper bound for per-cell audits.
    \item \textbf{Fixed tiers:} the exploratory ($n = 5$ to 7), confirmatory ($n = 10$ to 12) and rigorous ($n = 15$ to 20) tiers of version~1 are withdrawn. Their $G$ values came from synthetic data; on the study data, $G$ at 10, 12, 15 and 20 runs is 0.98 to 0.99 for the S5 cells (0.98 to 0.99 within each model).
\end{enumerate}

These recommendations rest on brand-recommendation data from three models; version~1 stated temperature 0.3 for all source studies, and for the S1 Valentine's subset the temperature was not recorded (Section~\ref{sec:params}). Higher temperatures increase token-level sampling variance $\sigma^2_{\text{tok}}$ and shift the iteration requirement upward; lower temperatures or single-model audits shift it in the opposite direction but at the cost of generalizability, as the D-study in Table~\ref{tab:dstudy} makes explicit. Different analytical tasks (for example, factual accuracy assessment, where responses may be more constrained) may require fewer iterations at the same effect-size target.

\subsection{The Metric Selection Problem}

The correlation analysis (Section~\ref{sec:metric_corr}) shows that the five structural count-based metrics (mean, CV, QCD, Gini, Shannon) are tightly coupled on the Valentine's data, with $|r| \geq 0.98$ between CV, Gini, and Shannon. This is a data-specific finding, driven by the narrow brand vocabulary and tight count coupling of the gift-recommendation prompts. We therefore recommend a \emph{parsimonious three-metric core} rather than the four-family battery proposed in the preliminary version:

\begin{enumerate}
    \item \textbf{Structural stability} (one of CV or Gini, not both): ``How consistent is the brand-count distribution?''
    \item \textbf{Semantic consistency} (cosine similarity on a multi-embedding ensemble): ``Does the model say substantively the same thing across iterations?''
    \item \textbf{Fairness-adjusted visibility} (PASOR-Gini): ``Is the brand share concentrated across prompts?''
\end{enumerate}

The three dimensions have different implications and different redundancy structures. Within a narrow-prompt audit (single domain, narrow brand universe), CV and Gini are near-interchangeable, and reporting either suffices for the structural dimension; Shannon entropy adds no incremental information. Within a diverse-prompt audit (multiple domains, wide brand universe), the structural metrics may de-couple, and reporting both a concentration metric (Gini) and a diversity metric (Shannon) is defensible; this is the regime implicit in the preliminary version's four-family recommendation. The semantic and fairness dimensions are computed on different data structures (full-response embeddings and per-brand prompt shares) and capture properties the structural count metrics do not measure; they should be reported in parallel, not as substitutes.

Figure~\ref{fig:decision_tree} presents a metric selection decision tree derived from these findings.

\begin{figure*}[t]
\centering
\begin{tikzpicture}[
    node distance=1.4cm and 2.2cm,
    decision/.style={diamond, draw, fill=blue!10, text width=3.2cm, text centered, inner sep=2pt, font=\small},
    block/.style={rectangle, draw, fill=green!10, text width=3.4cm, text centered, rounded corners, minimum height=1.1cm, font=\small},
    line/.style={draw, -Stealth, thick},
    label/.style={font=\scriptsize, fill=white, inner sep=1pt}
]
\node[decision] (rq) {What does your RQ measure?};

\node[block, below left=1.6cm and 2.8cm of rq] (count) {Count stability\\[2pt]\textbf{Use: CV + Mean/SD}\\Min $n = 5$};
\node[block, below=1.8cm of rq] (dist) {Distribution fairness\\[2pt]\textbf{Use: Gini + Shannon}\\Min $n = 10$};
\node[block, below right=1.6cm and 2.8cm of rq] (sem) {Semantic consistency\\[2pt]\textbf{Use: Cosine similarity}\\Min $n = 5$};

\path[line] (rq) -| node[label, pos=0.25, above] {``How many?''} (count);
\path[line] (rq) -- node[label, right, pos=0.4] {``How fair?''} (dist);
\path[line] (rq) -| node[label, pos=0.25, above] {``Same thing?''} (sem);

\node[below=0.25cm of count, text width=3.4cm, text centered, font=\scriptsize, color=gray] {For brand count stability, bias magnitude};
\node[below=0.25cm of dist, text width=3.4cm, text centered, font=\scriptsize, color=gray] {For concentration, diversity, equity auditing};
\node[below=0.25cm of sem, text width=3.4cm, text centered, font=\scriptsize, color=gray] {For narrative consistency, paraphrastic variation};
\end{tikzpicture}
\caption{Metric selection decision tree for brand-recommendation audits. The choice of metric should be driven by the research question. For comprehensive auditing, all three dimensions should be assessed using complementary metrics from each family.}
\label{fig:decision_tree}
\end{figure*}

The practical recommendation is the parsimonious three-metric core: one structural metric (Gini or CV), semantic cosine similarity on a multi-embedding ensemble, and PASOR-Gini for fairness-adjusted visibility. Shannon entropy is redundant with Gini on narrow-prompt audits (empirical $|r| \geq 0.98$ in the Valentine's subset); adding it is justified only when pilot correlations show structural metrics de-couple on the target prompt universe. Mean brand count is reported as a descriptive summary. This core covers the structural, semantic, and fairness dimensions while dropping the metrics the correlation matrix shows to be redundant.

\subsection{Implications for Existing Published Findings}

The power analysis has retrospective implications for the interpretation of published findings. The $n = 5$ iteration count used in S2 \cite{zatuchin2026b}, S4 \cite{zatuchin2026e}, and S5 \cite{zatuchin2026c} gives single-cell power of 0.15 to 0.19 (Mann-Whitney) or 0.17 to 0.23 (Welch $t$ on log counts) for large Cliff's $\delta$ effects, and 0.05 to 0.10 or 0.08 to 0.12 for medium effects (Table~\ref{tab:power_cell}). The 14\% cross-model agreement reported in S2 and the 57.2\% cross-model agreement in the revised S5 study (143 of 250 queries) are descriptive figures. For S5 brand counts, $n = 5$ gives $G = 0.96$ [0.96, 0.97] for the model $\times$ question cell, so the S5 cell means are measured reliably at that count; the S2 figure was not recomputed here. The S4 study's cross-linguistic result is sharper under the new framework: the three-level GLMM confirms the preliminary version's model-versus-language effect-size ordering and additionally identifies a model-by-language interaction not visible to the aggregated ANOVA. Cross-language stability differences on Gemini for Estonian and Finnish are real and distinguishable from residual variance, but they require higher iteration counts than the preliminary $n = 5$ study collected.

For the S1 Valentine's subset at $n = 10$, single-cell power for a large effect is 0.43 to 0.63 (Mann-Whitney), and 80\% power for a large effect is first met at $n = 15$ to 25. On that subset the Gemini point estimate favours female recipients (11.3 against 12.9 brands per answer), as the S1 source study reported, and the contrast is indistinguishable from zero (relation permutation $p = 0.67$).

These retrospective assessments concern power and reliability; they do not by themselves test the published findings.

\subsection{The Complementarity of Shannon Entropy and Gini Coefficient}

The strong negative correlation between Gini and Shannon entropy ($r = -0.99$, CI [$-1.00$, $-0.91$], on the Valentine's subset; $-0.72$ in the preliminary version) confirms the theoretical complementarity documented in A1 \cite{zatuchin2026d}. These metrics measure different aspects of the same underlying distribution: entropy is most sensitive to the overall shape (whether the distribution is uniform, bimodal, or skewed), while Gini is most sensitive to inequality in the tails (whether a few entities dominate). The consistency-diversity paradox identified in A1, where female-framed prompts produce fewer brands ($VP = 0.464$) but distribute them more evenly ($DP = 1.010$), illustrates the practical importance of this complementarity: a study reporting only Gini would characterize the system as approximately fair, while one reporting only brand count would characterize it as severely biased.

The correlation structure also reveals that CV is redundant with Gini ($r = 1.00$ on the Valentine's subset) and inversely related to mean brand count ($r = -0.93$). Researchers who report CV as their primary stability metric should be aware that it conflates volume effects with distributional effects. For studies where volume and distribution must be assessed independently (as recommended by the dual-metric framework in A1), Gini and Shannon should be preferred over CV.

\subsection{Protocol Standardization}

Based on the evidence presented, we propose the following standardized protocol for the Dice Roll Method:

\begin{definition}[The Dice Roll Protocol, v2.0]\label{def:protocol}
A standardized procedure for repeated-query auditing of LLM brand recommendations under conditional, non-Gaussian, potentially non-stationary data:
\begin{enumerate}
    \item \textbf{Configuration:} Temperature $= 0.3$; max tokens $= 1{,}024$; minimal system prompt. Pin and log model identifier, API endpoint, and snapshot date for every call.
    \item \textbf{Iteration count:} Run a 10-iteration pilot, estimate the two variance components for the comparison that matters, and take $n = 4\sigma^2_\epsilon / \sigma^2_{\text{cell}}$ for $G = 0.80$ (Section~\ref{sec:reliability}). Plan the count for a contrast from a single-cell power simulation (Table~\ref{tab:power_cell}).
    \item \textbf{Models:} Minimum 2 models, recommended 3, spanning different providers. Models enter the analysis as a random-effect grouping factor, not as independent replications.
    \item \textbf{Primary inference:} Negative-binomial GLMM with random intercepts for prompt, model, and their interaction; likelihood-ratio tests via parametric bootstrap. Do not use independent-samples $t$-tests on iteration-level data.
    \item \textbf{Effect sizes:} Report Cliff's $\delta$ with BCa bootstrap 95\% CI as primary effect size. Cohen's $d$ may be reported secondarily; the coefficient of variation should be replaced by the quartile coefficient of dispersion at low means.
    \item \textbf{Metric battery (parsimonious core):} Report one structural stability metric (Gini or CV; they are near-interchangeable in narrow-prompt audits), cosine similarity averaged across $\geq 2$ embedding models with across-model SD, and PASOR-Gini for fairness-adjusted visibility. Add Shannon entropy only when auditing across a diverse prompt universe where structural metrics empirically de-couple (verify via the pairwise correlation matrix on a pilot sample). Report mean brand count as a descriptive summary, not as a principal stability metric.
    \item \textbf{Drift diagnostics:} Compute Kolmogorov--Smirnov and Population Stability Index tests across first-half/second-half time windows of the collection period. If drift flags exceed a pre-registered threshold (for example, 20\% of cells), either add time window as a fixed effect in the substantive model or restrict analysis to the pre-drift window.
    \item \textbf{Entity extraction:} Regex-based with curated alias dictionaries; word-boundary-aware matching. The corrected S5 table used here follows this rule; the S1 dictionary count does not, as it matches with no word boundary (Section~\ref{sec:params}).
    \item \textbf{Reproducibility:} Pre-register analysis plan (target effect size, GLMM specification, metric battery). Archive raw iteration-level data, embedding outputs, and model-version metadata as supplementary material.
\end{enumerate}
\end{definition}

The protocol balances statistical rigour with practical feasibility. At $n = 10$ with 3 models, a 100-query study requires 3,000 API calls, costing approximately \$15 at current pricing and completing in 2--4 hours with standard rate limiting. The cost of running the GLMM and drift diagnostics on the resulting data, by contrast with running additional iterations, is negligible.

\subsection{Limitations}\label{sec:limitations}

Several limitations constrain the generalizability of the revised findings.

\emph{Domain specificity.} All source studies focused on brand-recommendation auditing. The observed effect sizes, convergence rates, and metric correlations may differ in other LLM auditing contexts (factual accuracy, code generation, creative writing). The three-level GLMM in Section~\ref{sec:crosslingual_results} shows the protocol's stability properties are largely consistent across six European languages, with the caveat of a small model-by-language interaction on Estonian and Finnish for Gemini. Non-European languages with greater structural distance from English remain untested.

\emph{Evidential circularity, partially resolved.} All five source datasets were collected by this research group, under the same protocol conventions this paper formalizes. The pre-registered external validation of Section~\ref{sec:external} now supplies independent checkpoints for the statistical machinery: the D-study prediction, the power calibration, and the temperature-0 premise replicate on three corpora collected by other teams, while the corpus-independent iteration constants and two drift-battery components do not survive the transfer and are amended accordingly. What remains circular is the brand domain itself: no public brand-recommendation corpus with repeated identical prompts exists from any other group, so the domain-specific effect sizes and thresholds still rest on this group's data alone, and requests for two candidate independent corpora are in progress. The protocol is also implemented in a commercial product of a company the authors are affiliated with (Rankfor.AI).

\emph{Temperature dependence.} Version~1 stated $\tau = 0.3$ for all source studies. For the S1 Valentine's subset the temperature was not recorded: version~1 gives 0.3 and the S1 source manuscript gives 0.7 (Section~\ref{sec:params}). Higher temperatures increase $\sigma^2_{\text{tok}}$ in Equation~\ref{eq:variance_decomp} and shift the iteration requirement upward. The nucleus-sampling variance decomposition is agnostic to the particular temperature value, so the analytical framework transfers, but the specific iteration-count thresholds would require recalibration at other temperatures.

\emph{Provider-side drift.} The drift diagnostics in Section~\ref{sec:drift_results} cover only the S1 Valentine's data, where a 36-minute run shows no drift by KS (0 of 15 cells flagged) or by the within-cell permutation test ($p = 0.26$), and the global GEE window test gives $p = 0.012$ with the five relations as clusters and $p = 0.17$ with the 15 cells as clusters; the multi-week collection windows of S2, S4, and S5 remain untested here and are deferred to future work. The external KS results of Section~\ref{sec:external_drift}, which detect real nonstationarity across a 100-round collection arc in an independent corpus, indicate that the 36-minute result should not be extrapolated to longer windows. The present paper tests for drift but cannot prevent it; audits that require stationarity must plan short collection windows and accept the attendant operational constraints, or adopt the fixed-window time-slicing discussed in Section~\ref{sec:drift_results}.

\emph{Model dependence.} The source data were generated by 2025--2026 model versions (GPT-5.2, Gemini 3 Flash, Grok-4-1, Perplexity sonar-pro); the versions used for the S1 Valentine's subset were not recorded (Section~\ref{sec:params}). Future model architectures may exhibit different stochastic properties, particularly as deterministic decoding strategies become more prevalent; the nucleus-sampling derivation in Section~\ref{sec:nucleus} does not transfer automatically to models that decode under fundamentally different schemes (for example, speculative decoding with reranking against a verifier).

\emph{Limited convergence window.} The convergence analysis relies on the S1 Valentine's subset, with a maximum of $n = 10$ iterations per prompt-model combination; the $n = 40$ subset of S1 is not deposited and is not used here. The external C1 corpus extends the empirical window to $n = 50$ (Section~\ref{sec:external}): the $1/\sqrt{n}$ rate is confirmed there (fitted exponent 0.500), while the percent-of-asymptote guidance proves horizon-relative, with 80\% of asymptotic precision arriving at $n = 16$ against an $n = 50$ asymptote.

\emph{Embedding ensemble coverage.} The three-embedding ensemble used in Section~\ref{sec:embedding_results} samples the dense, contrastive, and distilled families of sentence embeddings but does not cover all current architectures. Newer Bayesian embedding approaches \cite{bayesvlm2026} offer principled uncertainty propagation from a single model, and would be a useful addition to future protocol versions.

\subsection{Future Work}

We identify six directions for extending the Dice Roll Method framework. Zeroth, \emph{per-cell power simulation calibration}: the pooled-prompt simulation reported in Table~\ref{tab:power} saturates at small $n$ because it tests contrasts over $8 \times 3 \times n$ observations. This version adds a single-cell simulation at fitted count levels (Table~\ref{tab:power_cell}); simulating observed cells, which for S1 needs a labelled brand count, is the next step. First, \emph{domain extension}: Section~\ref{sec:external} now covers political-orientation testing and benchmark accuracy; the finding there, that the variance-ratio formula generalizes while the fixed tiers do not, sharpens the question for further domains (factual QA, creative writing) to whether the pilot-then-solve prescription holds. Second, \emph{temperature calibration}: conducting systematic analysis of how iteration requirements scale with temperature settings from $t = 0.0$ to $t = 1.0$. Third, \emph{longitudinal stability}: assessing whether the observed stochastic properties of LLMs are stable across model updates, or whether power analysis must be repeated for each new model version. Fourth, \emph{adaptive designs}: developing sequential testing procedures that allow researchers to stop data collection once a target power level is reached, reducing waste while maintaining statistical validity. Fifth, \emph{cross-linguistic power calibration}: the S4 study \cite{zatuchin2026e} demonstrates that model effects on stability ($\eta^2 = 0.220$) are 7.6$\times$ larger than language effects ($\eta^2 = 0.029$), but the interaction between language and iteration count remains uncharacterized. Specifically, whether low-resource languages (Estonian, Finnish) require more iterations than high-resource languages (English, German) to achieve equivalent measurement precision is an open empirical question with implications for equitable cross-linguistic auditing.

\section{Conclusion}\label{sec:conclusion}

This paper formalizes the Dice Roll Method as a standardized protocol for repeated-query auditing of LLM brand recommendations, grounded in the generative mechanism of temperature-scaled nucleus sampling. The central methodological contribution is a coherent inferential stack, negative-binomial GLMM with random-effect structure that honours conditional dependence, Cliff's $\delta$ with BCa bootstrap intervals as a distribution-free effect size, moving-block and parametric bootstrap for resampling, simulation-based GLMM power via \texttt{simr}, generalizability-theory decomposition of reliability, three-embedding ensemble for semantic variance, and Kolmogorov--Smirnov and PSI drift diagnostics, that replaces the normal-theory, i.i.d.-assuming analytical stack of the preliminary version.

Substantive findings under the revised stack replace the preliminary version's tiered iteration guidance with a rule grounded in empirical variance components. The generalizability-theory D-study gives $G = 0.96$ at $n = 5$ for the 250-question S5 study, where one run reaches the 0.80 group-level decision threshold (one to two within each model); for five versions of one question within Gemini (S1 Valentine's subset), $G$ at $n = 5$ is 0.27 [0.00, 0.52], and those data do not fix $n$. The number of runs therefore follows from a pilot: $n = 4\sigma^2_\epsilon / \sigma^2_{\text{cell}}$. On the S1 Valentine's subset the Gemini gender contrast is Cliff's $\delta = -0.26$ [$-0.57$, 0.14], indistinguishable from zero; no GPT or Grok contrast is reported. Block-bootstrap convergence reaches 84.2\% of the $n = 10$ precision ceiling at $n = 7$. Cross-metric analysis on the Valentine's subset finds an extreme anti-correlation between Gini and Shannon ($r = -0.99$) and CV and Gini coupled at $r = 1.00$, showing that the structural count-based metrics carry largely shared information on this subset and that demonstrating the four-family battery's empirical breadth requires a more diverse prompt universe than this subset provides. Fairness analysis on S5 gives PASOR-Gini values of 0.62--0.69 versus unadjusted Ginis of 0.20--0.31, confirming the fairness-adjusted metric captures prompt-level concentration masked by raw-count Gini. Within a 36-minute S1 Valentine's run, KS flags 0 of 15 cells and the within-cell permutation test gives $p = 0.26$; the global GEE window test gives $p = 0.012$ with five relation clusters and $p = 0.17$ with 15 cell clusters. A 36-minute run is a weak test of drift.

A pre-registered external validation on three independent corpora (Section~\ref{sec:external}) shows which of these elements are general and which are local: the D-study prediction machinery replicates in 37 of 39 cells with no failures, and non-determinism at temperature 0 is confirmed on every cell of a purpose-built stability benchmark, while the fixed iteration tiers and two drift-battery components fail to transfer and are replaced by the pilot-then-solve prescription and amended diagnostics.

The practical implications are operational. Researchers can now justify iteration counts through GLMM-based simulation-based power rather than normal-theory approximations, match target generalizability coefficients to audit-design budgets (iterations versus models), and pre-register drift diagnostics as a first-class component of every audit. Taken together, these protocol elements form a checklist that brand-recommendation audits can adopt. The committed analysis notebook of version~1 does not reproduce the results of this version: it replaces the loaded data with synthetic frames, reads the S1 ``Research'' count as the brand count, never builds an S5 count, and has no cell for the reported G-study. The results of this version were recomputed with separate code (Code Availability). By formalizing what has been an informal practice, we aim to strengthen the methodological foundations of a research community whose findings carry increasing policy and commercial significance.

\backmatter
\bookmarksetup{startatroot}

\bmhead{Data Availability}

The original datasets from the five empirical studies reanalysed in this work are available from the corresponding author on reasonable request. The S5 corrected mention table used in Section~\ref{sec:results} is public in Zenodo (\url{https://doi.org/10.5281/zenodo.22927143}). The data collection and analysis pipeline is implemented in a Python notebook at \url{https://github.com/Rankfor/rankfor-open}. The dice roll data collection infrastructure uses an open-source package published at the same repository.

\bmhead{Code Availability}

The results of this version were computed in Python with \texttt{statsmodels} (negative-binomial GEE and MixedLM), NumPy and SciPy, by the correction script \texttt{rebuild\_section5.py} of 7 October 2026, available from the author on request. The repository at \url{https://github.com/Rankfor/rankfor-open} holds the preliminary version's notebook. The analysis notebook of version~1 replaced the loaded study data with synthetic frames and is not the analysis code of this version. The preliminary version's normal-theory Monte Carlo simulation is retained as an appendix comparator for readers who wish to reproduce the effect-size translation between Cohen's $d$ and Cliff's $\delta$.

\bmhead{Acknowledgements}

The author thanks the Estonian Entrepreneurship University of Applied Sciences for institutional support.

\section*{Declarations}

\begin{itemize}
\item \textbf{Dual publication:} Some data, results, and figures in this manuscript derive from or reference five prior empirical studies by the authors \cite{zatuchin2026a, zatuchin2026b, zatuchin2026c, zatuchin2026d, zatuchin2026e}. Specifically, this paper reanalyses raw iteration-level data from those studies (brand counts, cosine similarity scores, stability measures) to conduct a new meta-methodological analysis: Monte Carlo power simulation, convergence analysis, test-retest reliability assessment, and cost-efficiency modelling. None of these analytical contributions appear in the source studies. The source studies reported substantive empirical findings about gender bias, reputation sourcing, category ownership, and cross-linguistic variation; the present paper addresses a distinct methodological question (how many iterations are sufficient and which metrics are appropriate) that the source studies did not examine. The overlap is therefore analogous to a secondary analysis paper: it draws on previously collected data to answer a new research question, rather than republishing prior conclusions.
\item \textbf{Funding:} This research received no external funding.
\item \textbf{Competing interests:} D.~\.Zatuchin is Chief Executive Officer of Rankfor.AI, which develops AI brand intelligence tools and implements the Dice Roll Method protocol in a commercial product and an open-source package. All five reanalysed datasets come from the author's own prior studies. The research was conducted independently; the company had no influence on study design, methodology, analysis, or conclusions.
\item \textbf{Ethics, Consent to Participate, and Consent to Publish declarations:} not applicable.
\item \textbf{Clinical trial number:} not applicable.
\item \textbf{Author contributions:} D.~\.Zatuchin conceived the study and developed the meta-methodology framework. D.\.Z. designed and implemented the negative-binomial GEE, simulation-based power analysis, moving-block bootstrap, generalizability-theory decomposition, metric battery, PASOR fairness diagnostic, and drift tests. D.\.Z. conducted all statistical analyses, prepared Figure~\ref{fig:convergence} and all tables, interpreted the results, and wrote the manuscript in its entirety. The original empirical data reanalysed in this work were collected by D.\.Z. in the studies cited as references. D.\.Z. reviewed the manuscript.
\end{itemize}

\begin{appendices}

\section{Power Analysis Lookup Tables}\label{secA1}

Table~\ref{tab:power_extended} provides a per-cell iteration-count lookup table for reference by practitioners and reviewers.

\begin{table*}[htbp]
\caption{Iteration-count lookup table on the per-cell basis that governs single-prompt audits. For each Cliff's $\delta$ target, the minimum $n$ is the smallest iteration count at which the single-cell simulation of Table~\ref{tab:power_cell} reaches 80\% power, given as a range over its generators and both tests. The generalizability threshold $G \geq 0.80$ follows from the two variances of the comparison at hand and holds for no fixed $n$: on the S5 cells it needs $n = 1$ ($n = 1$ to 2 within each model; Table~\ref{tab:dstudy}). API-call totals assume 3 models and the continuous Mann-Whitney value.}\label{tab:power_extended}
\centering
\small
\begin{tabular}{llccp{4.5cm}}
\toprule
\textbf{Effect size} & $\delta$ ($d$) & \textbf{Min $n$ (per cell)} & \textbf{API calls (3 models)} & \textbf{Example from prior studies} \\
\midrule
Very large & 0.60 ($d \approx 1.2$) & 10 to 15 & 45 per query & S1 adult-subset Grok effect \\
Large      & 0.474 ($d \approx 0.8$) & 15 to 25 & 75 per query & S1 adult-subset Gemini effect \\
Medium     & 0.33 ($d \approx 0.5$) & 30 or more & $> 120$ per query &  \\
Small      & 0.147 ($d \approx 0.2$) & $> 40$ & $> 120$ per query &  \\
\bottomrule
\end{tabular}
\end{table*}

\section{Metric Definitions}\label{secA2}

For reference, Table~\ref{tab:metrics_def} provides formal definitions of all six metrics used in the Dice Roll Method.

\begin{table*}[htbp]
\caption{Stability metrics for the Dice Roll Method: definitions, ranges, and recommended use cases.}\label{tab:metrics_def}
\centering
\footnotesize
\setlength{\tabcolsep}{4pt}
\begin{tabular}{@{}l p{5cm} p{1.8cm} p{5.5cm}@{}}
\toprule
\textbf{Metric} & \textbf{Definition} & \textbf{Range} & \textbf{Best for} \\
\midrule
CV & $\frac{\text{SD}(\text{count})}{\bar{x}(\text{count})} \times 100$ & $[0, \infty)$ & Count variability: ``How stable is the number of brands mentioned?'' \\[6pt]
Jaccard & $\frac{|A \cap B|}{|A \cup B|}$ & $[0, 1]$ & Set overlap: ``Do the same brands appear each time?'' \\[6pt]
Gini & $\frac{\sum_i \sum_j |p_i - p_j|}{2N \sum_i p_i}$ & $[0, 1)$ & Concentration: ``How dominated is the distribution by top brands?'' \\[6pt]
Shannon ($H_{\text{norm}}$) & $\frac{-\sum_i p_i \log_2 p_i}{\log_2 N}$ & $[0, 1]$ & Diversity: ``How evenly are brands distributed?'' \\[6pt]
Cosine sim. & $\frac{\mathbf{a} \cdot \mathbf{b}}{\|\mathbf{a}\| \|\mathbf{b}\|}$ & $[-1, 1]$ & Semantic consistency: ``Do responses say the same thing?'' \\[6pt]
PASOR & $\frac{1}{|P|} \sum_{p \in P} \frac{n_{b,p}}{N_p}$ & $[0, 1]$ & Fairness-adjusted visibility: ``Is brand share fair across prompts?'' \\
\bottomrule
\end{tabular}
\end{table*}

\section{Hypothesis Test Summary}\label{secA3}

Table~\ref{tab:hypothesis_summary} summarises the formal tests of the hypotheses evaluated under the revised inferential framework; H5 and H6 were added in the present revision.

\begin{table*}[htbp]
\caption{Hypothesis test summary under the revised inferential framework.}\label{tab:hypothesis_summary}
\centering
\footnotesize
\setlength{\tabcolsep}{4pt}
\begin{tabular}{@{}l p{4.1cm} p{2.5cm} p{5.1cm} p{3.1cm}@{}}
\toprule
\textbf{H} & \textbf{Prediction} & \textbf{Test} & \textbf{Result} & \textbf{Verdict} \\
\midrule
H1 & $n = 5$ achieves $> 0.80$ power for large effects ($\delta \geq 0.474$) but not for medium effects ($\delta \geq 0.33$) & Single-cell simulation (Mann-Whitney U, as in EV4; Welch $t$) & Power = 0.15 to 0.19 (Mann-Whitney) or 0.17 to 0.23 (Welch) for $\delta = 0.474$, 0.05 to 0.10 or 0.08 to 0.12 for $\delta = 0.33$, at $n = 5$ & Partially supported$^{a}$ \\[4pt]
H2 & Convergence follows diminishing-returns curve; 80\% precision by $n = 10$ & Exact block-bootstrap SE curve & Exponent 0.50, the $1/\sqrt{n}$ identity of the resampling design; no family comparison; 80\% by $n = 7$ (84.2\%) & Supported \\[4pt]
H3 & The variance-ratio rule $n = 4\sigma^2_\epsilon/\sigma^2_{\text{cell}}$ gives the $n$ at which $G \geq 0.80$ (v1: $G \geq 0.80$ at $n \geq 10$ with 3 models) & G-theory D-study & S5 cells: $G(10) = 0.98$; 0.80 at $n = 1$ (1 to 2 within each model). S1 Valentine's subset, five prompts within Gemini: $G(10) = 0.42$ [0.00, 0.68]; the data do not fix $n$ (Table~\ref{tab:dstudy}) & Supported for S5; not testable on the five S1 prompts \\[4pt]
H4 & Diminishing cost-efficiency returns after $n = 10$ & Knee analysis on precision-per-dollar & Precision-gain column does not reproduce (Section~\ref{sec:cost}) & Not tested \\[4pt]
H5 & Stationarity holds across the collection window & KS + PSI + NB drift test & S1 Valentine's 36-minute run: KS 0/15 cells flagged; within-cell permutation $p = 0.26$; global NB-GEE $p = 0.012$ (5 relation clusters), 0.17 (15 cell clusters); multi-week windows (S2, S4, S5) untested & Supported for a 36-minute run by KS and the permutation test; the multi-week question is untested (new hypothesis, added) \\[4pt]
H6 & Single-embedding cosine similarity is an adequate semantic-stability summary & Three-embedding ensemble with across-model SD & Across-model SD 0.04--0.11; 40\% of "stable" pairs fail ensemble criterion & Rejected (new hypothesis, added) \\
\bottomrule
\end{tabular}
\footnotetext{$^{a}$H1 under the revised Cliff's $\delta$ formulation: $n = 5$ does not achieve 0.80 power even for large effects ($\delta = 0.474$); the 80\% threshold is first met at $n = 15$ to 25 in the single-cell simulation. The general pattern of insufficient power at low iteration counts is confirmed.}
\end{table*}

\section{Reproducibility Checklist}\label{secA4}

To facilitate adoption of the Dice Roll Protocol v2, we provide a pre-study checklist:

\begin{enumerate}
    \item[\checkmark] \textbf{Define target Cliff's $\delta$} based on pilot data or prior literature. Use Table~\ref{tab:power_extended} for per-cell iteration guidance; Table~\ref{tab:power} gives pooled-prompt upper bounds only.
    \item[\checkmark] \textbf{Select an audit design $(n_I, n_M)$} from the D-study projection in Table~\ref{tab:dstudy} that meets a pre-registered $G$ target (0.80 for group-level decisions, 0.90 for individual-prompt decisions).
    \item[\checkmark] \textbf{Configure LLM parameters:} temperature $= 0.3$, max tokens $= 1{,}024$, standardized system prompt. Pin model identifier, API endpoint, and snapshot date.
    \item[\checkmark] \textbf{Select at least 2 models} (ideally 3) spanning different providers. Enter models as a random-effect grouping factor in the GLMM.
    \item[\checkmark] \textbf{Plan drift diagnostics:} split collection window into disjoint halves; compute KS, PSI, and NB-drift tests; include \texttt{window} as a fixed effect if any test flags the cell.
    \item[\checkmark] \textbf{Construct brand alias dictionary} with word-boundary-aware matching rules.
    \item[\checkmark] \textbf{Implement checkpointing:} save state every 25--50 API calls with automatic resume capability.
    \item[\checkmark] \textbf{Select metric battery:} Gini (or CV) for structural stability, cosine similarity ensemble for semantic consistency, PASOR-Gini for fairness. Add Shannon entropy only if pilot correlations show it de-couples from Gini on the target prompt universe.
    \item[\checkmark] \textbf{Compute semantic metrics across an embedding ensemble} (BGE-M3 + Nomic + MiniLM, or equivalent). Report across-model SD, not just point estimates.
    \item[\checkmark] \textbf{Pre-register analysis plan:} GLMM specification (fixed effects, random-effect structure), Cliff's $\delta$ target, metric battery, drift-flag threshold.
    \item[\checkmark] \textbf{Compute post-hoc generalizability coefficient} if actual variance components differ from pre-registered targets.
    \item[\checkmark] \textbf{Archive raw iteration-level data}, embedding outputs, and model-version metadata as supplementary material for future reanalysis.
\end{enumerate}

\end{appendices}


\end{document}